\documentclass[aps,prd,twocolumn,
               notitlepage,mathrsfs,
               amssymb,amsmath,amsfonts,
               superscriptaddress,
               longbibliography,
               nofootinbib,floatfix]{revtex4-2}
              
\usepackage[normalem]{ulem}
\usepackage{graphicx}
\usepackage[linktocpage,breaklinks]{hyperref}
\usepackage[capitalize]{cleveref}
\usepackage[usenames,dvipsnames]{xcolor}
\hypersetup{colorlinks=true,
            citecolor=NavyBlue,
            linkcolor=magenta,
            urlcolor=magenta}
\usepackage{multirow,array}
\DeclareMathAlphabet{\pazocal}{OMS}{zplm}{m}{n}
\usepackage{journals}
\usepackage{enumerate}
\usepackage[caption=false]{subfig}

\newcommand{\Tns}{\tau_{\rm H}}
\newcommand{\Tsc}{\tau_{\rm S}}
\newcommand{\comp}{\lambdabar_{\rm comp}}
\newcommand{\pa}{\partial}

\newcommand{\addal}[1]{{\color{blue} #1}} % added by Alan

\begin{document}
%%%%%%%%%%%%%%%%%%%%%%%%%%%%%%%%%%
% % %
\title{Binary neutron star mergers in massive scalar-tensor theory: Properties of post-merger remnants}

\date{\today}
%%  Add Abbreviations for Journals

\author{Alan Tsz-Lok Lam}
\email{tszlok.lam@aei.mpg.de}
\affiliation{Max Planck Institute for Gravitational Physics (Albert Einstein Institute), 14476 Potsdam, Germany}

\author{Hao-Jui Kuan}
\affiliation{Max Planck Institute for Gravitational Physics (Albert Einstein Institute), 14476 Potsdam, Germany}

\author{Masaru Shibata}
\affiliation{Max Planck Institute for Gravitational Physics (Albert Einstein Institute), 14476 Potsdam, Germany}
\affiliation{Center of Gravitational Physics and Quantum Information, Yukawa Institute for Theoretical Physics, Kyoto University, Kyoto, 606-8502, Japan} 

\author{Karim Van Aelst}
%\email{karim.vanaelst@aei.mpg.de}
\affiliation{Max Planck Institute for Gravitational Physics (Albert Einstein Institute), 14476 Potsdam, Germany}

\author{Kenta Kiuchi}
\affiliation{Max Planck Institute for Gravitational Physics (Albert Einstein Institute), 14476 Potsdam, Germany}
\affiliation{Center of Gravitational Physics and Quantum Information, Yukawa Institute for Theoretical Physics, Kyoto University, Kyoto, 606-8502, Japan} 

\begin{abstract}
We investigate the properties of post-merger remnants of binary neutron star mergers in the framework of Damour-Esposito-Farese-type scalar-tensor theory of gravity with a massive scalar field by numerical relativity simulation.
It is found that the threshold mass for prompt collapse is raised in the presence of the excited scalar field.
Our simulation results also suggest the existence of long-lived $\phi-$mode in hypermassive neutron stars due to the presence of the massive scalar field which enhances the quasi-radial oscillation in the remnant.
We investigate the descalarization condition in hypermassive neutron stars and discover a distinctive signature in post-merger gravitational waves.
\end{abstract}
\maketitle

%%%%%%%%%%%%%%%%%%%%%%%%%%%%%%%%%%
\section{Introduction}
%%%%%%%%%%%%%%%%%%%%%%%%%%%%%%%%%%

After the monumental event GW170817 \cite{abbo17,abbo17b,abbo17c,kasl17}, huge effort has been devoted to modeling the physics involved in the course of binary neutron star (BNS) mergers with the hope of learning more about the nuclear equation of state (EOS) of matters in extreme environment, exploring $r$-process nucleosynthesis in the merger ejecta, and understanding the non-linear nature of gravity. In particular, through measuring the size of matter effects of the neutron star (NS) members in the late inspiral stages for this event, the stiffness of the EOS has been constrained to a narrow range \cite{abbo18,de18,anna18,capa20,diet20}. In addition, general relativity (GR) has proven to accurately reproduce gravitational effects at least up to the stage shortly before the merger. Considering the Damour-Esposito-Farese type  extension to GR (DEF theory in what follows), this can be translated to an upper bound on the coupling constant, which prohibits spontaneous scalarization in isolated NSs for massless scalar field \cite{zhao19} while admitting of mild scalarization for massive cases \cite{kuan23}.
A plausible agent to push the known constraints further is the remnant system in the aftermath of the merger, where higher-energy physics, for which details have not been yet understood, can play an important role. The evolution process of BNS remnants is also the key determinant of multimessenger signals~\cite{shib19,bern20}: the properties of the electromagnetic (EM) signals depend strongly on the mass and the composition of ejecta from the remnant including some ultra-relativistic jets~\cite{kawa19}, and post-merger gravitational waves (GWs) encode information about BNS parameters~\cite{shib05b,kiuc09,hoto13}.

Joint detection of EM and GW signals provides a unique avenue to learn the details of post-merger systems such as the lifetime of the remnant NSs. The latter quantity is sensitive to the EOS and underlying gravitational theory. Although GR functions quite well throughout the inspiral history of binaries, beyond-GR signatures may reveal shortly before, during, and after the merger. For example, the DEF theory can admit dynamical scalarization and/or enhanced scalar cloud in the parameter region corresponding to GW170817 \cite{kuan23}. Besides, the additional scalar degree of freedom can lead to qualitative differences in the post-merger waveform, and impact the evolution of the object produced in the merger. The goal here is thus to extensively investigate the outcomes of BNS mergers in the DEF theory, whereas magnetic, neutrino, and thermal physics are not taken into account as we focus on the post-merger stage only for a short timescale. 

In most BNS mergers, either a hypermassive neutron star (HMNS, which is stabilized by a high degree of angular momentum with a differential rotation \cite{shib99,baum00,shib00}) is formed and lives for some time before collapsing to a black hole, or a prompt collapse occurs if the total mass of the BNS exceeds a threshold $M_{\rm thr}$. The threshold mass for the prompt collapse is sensitive to the nuclear EOS~\cite{shib06,shib19}. On the other hand, it is expected to be rare that a supramassive NS is produced from a BNS since the total mass of the system should be less than the maximum mass that is supportable by rigid rotation ($M_c$). An empirical relation of such critical value is $M_c\simeq1.2M_{\rm TOV}$ with $M_{\rm TOV}$ the maximum mass of a spherical cold NS of a given EOS \cite{cook92,cook94,cook94b}, which then suggests $M_c\alt2.6\,M_\odot$ (e.g.,~\cite{shib17,rezz18}). Some population studies thus suggest that only $\alt15\%$ of BNSs has a total mass lower than $M_c$ \cite{farr19} (see also \cite{taur17}). In the present work, we focus on scenarios with total mass larger than $M_c$, i.e., a black hole + torus will be formed either shortly after the merger or after the rotational profile is modified within the HMNS~\cite{duez04,hoto13c}.

The presence of a torus surrounding the black hole plays an essential role in determining the post-merger emissions, such as short gamma ray bursts~\cite{shib06,kiuc10} and kilonovae~\cite{fuji20,kawa23,curt23,comb23}. The amount of matter ejected to form the torus depends strongly on the total mass, and the nuclear EOS for both prompt collapse and HMNS formation scenarios~\cite{kiuc09,hoto11,hoto13b,diet17} (see also \cite{shib19} for a review). In the latter scenario, the lifetime of HMNS, $\Tns$, is the main factor that determines the torus mass especially when the BNS is of (nearly) equal mass, since the matter injection from the central object ceases upon the formation of the black hole \cite{hoto13c}. 

It has been known that the value of $\Tns$ for short-lived HMNSs is determined primarily by the BNS's total mass if the system is moderately symmetric (e.g.,~\cite{shib06,hoto11,baus13,sven19,bern20,kash22}) in GR. 
Under the framework of the DEF theory, the lifetime of HMNSs is also likely to be sensitive to the scalar parameters, which are the strength of the coupling ($B$) of the scalar field to the metric functions, and the mass of the scalar field ($m_\phi$). In addition to their lifetime, the scalar field can also exist in the HMNSs for a certain time,  $\Tsc(\le\Tns)$.
Depending on $\Tns$, three possibilities for the outcome are generically expected: (i) prompt collapse to a black hole, (ii) short-lived HMNS formation, and (iii) long but finite lived HMNS formation. 
In the presence of an excited scalar field in the DEF theory, $\Tsc$ further divides channel (iii) into (iii.a) long-lived scalarized HMNSs and (iii.b) those descalarizing at some point. The two characteristic time-scales are dependent on the source and theory parameters, namely, the total mass and mass ratio of the BNSs, ($M_{\rm tot}$, $q$), the EOS, $B$, and $m_\phi$. The main goal of the present study is to  investigate how the two crucial timescales are modified by the scalar quantities by performing numerical-relativity simulations for equal-mass BNSs.

The paper is organized as follows. \cref{sec:formalism} briefly introduces the DEF theory, the associated $3+1$ decomposition for numerical evolution, the EOS employed, the details of the numerical setup, and the parameters we consider in this work.
In \cref{sec:postmerger} we discuss in detail the post-merger scenarios including the formation of a long-lived HMNS, a short-lived HMNS, and prompt collapse to a black hole, and investigate the effect of the scalar field on the HMNS lifetime and the threshold mass.
The properties of the remnant including dynamical ejecta, GW signal, mass of the final black hole and disk with potential quantities relevant to observation are given in \cref{sec:properties}. \cref{sec5} is devoted to summary and discussion.
Throughout this paper, we employ the geometrical units $c=G_*=\hbar=1$, where $c$, $G_*$, and $\hbar$ are the speed of light, the ``bare'' gravitational constant, and the reduced Planck constant, respectively. In the DEF theory, the gravitational constant in the action is $G=G_*\phi$. Subscripts $a,b,c,\dots$ running from 0 to 3 denote the spacetime components while $i, j, k, \dots$ running from 1 to 3 denote the spatial components. 

\section{Formalism} \label{sec:formalism}

In the so-called Jordan frame, the action of the DEF theory reads \cite{damo92}
\begin{align}
    S =& \frac{1}{16\pi}\int d^4x \sqrt{-g}
    \left[ \phi {\cal R} - \frac{\omega(\phi)}{\phi}
    \nabla_a \phi \nabla^a \phi - U(\phi) \right] \nonumber\\
    &- S_{\rm matter},
    %&- \int d^4x \sqrt{-g} \rho (1 + \epsilon),
\end{align}
where ${\cal R}$ is the Ricci scalar associated with metric $g_{ab}$, $\phi$ is the scalar field, $S_{\rm matter}$ is the action for matter, and $\omega(\phi)$ is chosen to have the form \cite{shib14,tani15}
\begin{align}
    \frac{1}{\omega(\phi)+3/2}= B\, \ln \phi,
\end{align}
for a coupling constant $B$. For the latter use, we introduce the auxiliary variable $\varphi$,
\begin{align}\label{eq:defvarphi}
    2\ln\phi = \varphi^2,
\end{align}
with respect to which the scalar potential is defined as \cite{kuro23,kuan23}
\begin{align}\label{eq:potential}
    U(\phi)=\frac{2m_\phi^2\varphi^2\phi^2}{B},
\end{align}
where it can be seen that $m_\phi$ is the scalar mass when one rewrites the potential into the so-called Einstein frame.
We assume that the asymptotic value of the scalar field, $\varphi_0$, vanishes at the spacial infinity, same as \cite{kuro23,kuan23}.

\subsection{Evolutionary Equations}

The associated equations for the metric and scalar fields can be derived as~(e.g., \cite{kuan23})
\begin{subequations}
\begin{align}\label{eq:einstein}
    G_{ab} &= 8\pi \phi^{-1} T_{ab} 
    +\omega(\phi)\phi^{-2} \left[ (\nabla_a\phi) (\nabla_b \phi) - \frac{1}{2} g_{ab} \nabla_c \phi \nabla^c \phi \right]\nonumber \\
    &+\phi^{-1} \left(\nabla_a\nabla_b \phi - g_{ab} \Box_g \phi\right) - \frac{m_\phi^2\varphi^2\phi}{B}g_{ab},
\end{align}
and
\begin{align}\label{eq:phi}
    \nabla_a \nabla^a \phi &= \frac{1}{2\omega(\phi) +3} \left[8\pi T - \frac{d \omega}{d\phi}
    (\nabla_c\phi)(\nabla^c\phi)+\frac{4m_\phi^2\phi^2}{B}\right] \nonumber\\
    &=2\pi\varphi^2 BT + \phi^{-1}\varphi^{-2}(\nabla_c\phi)(\nabla^c\phi) +m_\phi^2\varphi^2\phi^2,
\end{align}
\end{subequations}
where $G_{ab}$ and $\nabla_a$ are the Einstein tensor and covariant derivative associated with $g_{ab}$, and $T_{ab}$ is the stress-energy tensor with $T:=T_a^{~a}$. Since we evolve $\varphi$ rather than $\phi$, we rewrite \cref{eq:phi} in terms of $\varphi$ as
\begin{align}\label{eq:s_eq}
    \nabla_a \nabla^a \varphi =& 2\pi \phi^{-1} B T \varphi -\varphi(\nabla_c\varphi)(\nabla^c\varphi)+m_\phi^{2}\varphi\phi,
\end{align}
which will be used to derive the evolution equation for the auxiliary scalar field.

The evolution equations for gravitational and scalar fields can be derived by 3+1 decomposition (see Ref.~\cite{shib14} for the detailed derivation in the massless DEF case). Following the Baumgarte-Shapiro-Shibata-Nakamura (BSSN) formalism \cite{shib95,baug98}, we can obtain the modified evolution equations in the Cartesian coordinates as follows \cite{kuro23}:
%%%%%%%%%%%%%%%%
\begin{align}
    ( \pa_t - &\beta^k \pa_k ) W =
    \frac{1}{3} W \left( \alpha K - \pa_k \beta^k \right), \\
%%%
    ( \pa_t - &\beta^k \pa_k ) \tilde \gamma_{ij} = 
    - 2 \alpha \tilde A_{ij} \nonumber \\
    &+ \tilde \gamma_{ik} \pa_j \beta^k
    + \tilde \gamma_{jk} \pa_i \beta^k - \frac{2}{3} \tilde \gamma_{ij} \pa_k \beta^k, \\
%%%
    ( \pa_t - &\beta^k \pa_k ) \tilde A_{ij} = 
    W^2 \left[ \alpha R_{ij} -  D_i D_j \alpha - 8\pi \alpha \phi^{-1} S_{ij} \right]^{\rm TF} \nonumber \\
    &+ \alpha \left( K \tilde A_{ij} - 2 \tilde A_{ik} \tilde A_j{}^k \right)
    + \tilde A_{kj} \pa_i \beta^k + \tilde A_{ki} \pa_j \beta^k \nonumber\\
    &- \frac{2}{3} \tilde A_{ij} \pa_k \beta^k + \alpha \tilde A_{ij} \varphi \Phi \nonumber\\
    & - \alpha W^2 \left[ \omega \varphi^2 D_i \varphi D_j \varphi + \phi^{-1} D_i D_j \phi \right]^{\rm TF}, 
%%%
\end{align}
\begin{align}
    (\partial_t - &\beta^k \partial_k) K
    =\,\, 4\pi \alpha \phi^{-1} (S^i{}_i+\rho_{\rm h})+\alpha K_{ij} K^{ij}-D_i D^i \alpha \nonumber\\
    &+\alpha \omega \varphi^{2} \Phi^2 - \left( \frac{3}{2}+\frac{1}{B} \right)\alpha m_\phi^2\varphi^2\phi
    \nonumber \\
    &+ \alpha\phi^{-1}\Big[
    D_iD^i\phi-K\Phi\phi\varphi \nonumber\\
    &-3\pi\varphi^2 BT
    +\frac{3}{2\varphi^2\phi} \left(\Phi^2\phi^2\varphi^2-D_k\phi D^k\phi\right)
    \Big], \label{eq:K} \\
%%%
    (\pa_t - &\beta^k \pa_k ) \tilde \Gamma^i =
    2 \alpha \left( \tilde \Gamma^{i}_{jk} \tilde A^{jk} - \frac{2}{3} \tilde \gamma^{ij} \pa_j K 
    - \frac{3}{W} \tilde A^{ij} \pa_j W \right) \nonumber \\
    & - 2 \tilde A^{ij} \pa_j \alpha - 2\alpha \tilde \gamma^{ij} \left[ 8\pi \phi^{-1} J_j - \varphi K_j{}^k D_k\varphi \nonumber \right. \\
    & \left.+\left(1+ \frac{2}{B}-\frac{\varphi^2}{2} \right)\Phi D_j\varphi
    +\varphi D_j\Phi \right] \nonumber \\
    &+ \tilde \gamma^{jk} \pa_j \pa_k \beta^i + \frac{1}{3} \tilde \gamma^{ij} \pa_j \pa_k \beta^k \nonumber \\
    &- \tilde \gamma^{kl} \tilde \Gamma^j{}_{kl} \pa_j \beta^i + \frac{2}{3} \tilde \gamma^{jk} \tilde\Gamma^i{}_{jk} \pa_l \beta^l, 
\end{align}
\begin{align}
    (\partial_t -&\beta^k \partial_k)\varphi =-\alpha \Phi,\\
%%%
    (\partial_t -&\beta^k \partial_k)\Phi = -\alpha D_i D^i \varphi
    - (D_i \alpha)D^i\varphi
    -\alpha \varphi (\nabla_a \varphi)\nabla^a \varphi \nonumber\\
    &+\alpha K \Phi
    + 2\pi \alpha \phi^{-1} B T \varphi +\alpha m_\phi^2\varphi\phi,
%%%
\end{align}
where $\alpha$ is the lapse function, 
$\beta^i$ is the shift vector, 
$\Phi := - n^a \nabla_a \varphi$ is the "momentum" of the scalar field with $n^a=(1/\alpha, - \beta^i/\alpha)$, 
$\gamma_{ij}$ is the spatial metric with $\gamma := \det{(\gamma_{ij})}$, 
$W:= \gamma^{-1/6}$, 
$\tilde\gamma_{ij} := \gamma^{-1/3} \gamma_{ij}$ is the conformal spatial metric, 
$\tilde \Gamma^i{}_{jk}$ is the Christoffel symbol of $\tilde \gamma_{ij}$ with $\tilde \Gamma^i := - \pa_j \tilde\gamma^{ij}$, $(S_{ij})^{\rm TF}:= S_{ij} - \gamma_{ij} S^k{}_k / 3$ denotes the trace-free part of the stress tensor $S_{ij}$,  $K_{ij}$ is the extrinsic curvature with $K:=K^i{}_i$ being its trace, $\tilde A_{ij} := W^2 (K_{ij})^{\rm TF}$ is the conformal traceless part of $K_{ij}$, $S_{ij}:= \gamma^a{}_i \gamma^b{}_j T_{ab}$, $\rho_h := n^a n^b T_{ab}$, $J_i := - \gamma^a{}_i n^b T_{ab}$ are the spacetime decomposition of the stress-energy tensor, $R_{ij}$ is the spatial Ricci tensor, and $D_i$ is the covariant derivative with respect to the spatial metric.
We adopt the moving-puncture gauge \cite{camp06,bake06} for the lapse function and shift vector in the form:
\begin{align}
    ( \partial_t - \beta^j \partial_j ) \alpha &= - 2 \alpha K , \\
    ( \partial_t - \beta^j \partial_j ) \beta^i &= (3/4) B^i , \\
    ( \partial_t - \beta^j \partial_j ) B^i &= ( \partial_t - \beta^j
    \partial_j ) \tilde{\Gamma}^i - \eta_B B^i,
\end{align}
where $B^i$ is an auxiliary variable and $\eta_B$ is a parameter typically set to be $\sim 1/M_\mathrm{tot}$.
The Hamiltonian and momentum constraints can be found in Eqs.~(15) and (16) of \cite{kuan23}, and will not be repeated here.

In the Jordan frame, the scalar field does not affect the matter evolution explicitly, and thus, the equations of motion for matter are the same as those in GR. We assume a perfect fluid [i.e., $S_{\rm matter} = \int d^4 x \sqrt{-g} \rho ( 1 + \epsilon)$], for which the stress-energy tensor is expressed as
\begin{align}
    T^{ab} = \rho h u^a u^b + P g^{ab}, 
\end{align}
and the conservation equations are given by
\begin{align}
    \nabla_a T^a{}_b=0.\label{eq:tab}
\end{align}
Here $\rho$ is the rest-mass density, $P$ is the pressure, $\epsilon$ is the specific internal energy, $h:= 1 + \epsilon + P/\rho$ is the specific enthalpy, and $u^a$ is the four-velocity of the fluid. In addition to \cref{eq:tab}, we solve the continuity equation, $\nabla_a (\rho u^a)=0$.

\subsection{Equation of state}

We adopt the piecewise-polytropic approximation \cite{read09} for the barotropic EOS APR4 \cite{apr4}, MPA1 \cite{muth87}, and H4 \cite{lack06},
which cover a range of stiffness favored by GW170817 \cite{abbo18,fatt18,diet20}.
In addition, we adopt the following description for the thermal pressure, which is associated with the generation of shocks in the plunge and post-merger stages:
\begin{align}
    P = P_{\rm cold}(\rho) + P_{\rm th}(\rho,\epsilon),
\end{align}
where the cold contribution to the pressure, $P_{\rm cold}(\rho)$, is dictated by the cold EOS, and the thermal contribution is assumed to take the form \cite{jank93}
\begin{align}
    P_{\rm th}=(\Gamma_{\rm th}-1)\rho\epsilon_{\rm th},
\end{align}
with the adiabatic index $\Gamma_{\rm th}$ for heated matter, and $\epsilon_{\rm th}=\epsilon-\epsilon_{\rm cold}$ is the residual in the specific internal energy that is not included in the cold EOS. 
In general, $\Gamma_{\rm th}$ depends on the temperature and rest mass density~\cite{baus10}, while it has been suggested that a (reasonable) constant approximation suffices for investigating the fate of the merger remnant~\cite{shib05c,hoto11,baus13}. We choose $\Gamma_{\rm th}=1.8$ for our simulations.
Depending on the EOS and theory parameters, NSs in a coalescing BNS can remain unscalarized up to merger, be dynamically scalarized in the late inspiral, or be spontaneously scalarized at large separation \cite{bara13,pale14,shib14,tani15,senn16}. 

\subsection{Numerical setup}
We implement the Z4c version of the evolution equations by extending the code developed in \cite{yama08}, which was parallellised to \texttt{SACRA-MPI} in \cite{kiuc17}. 
\texttt{SACRA-MPI} employs a box-in-box adaptive mesh refinement with 2:1 refinement and imposes equatorial mirror symmetry on the $z=0$ orbital plane.
For the simulations shown in this article, each NS is covered by 4 comoving finer concentric boxes, with 6 coarser domains underneath containing both piles of the finer domains.
The size of the finest domain is chosen to be about $1.3$ to $1.5$ times of NS radius.
All domains are covered by $(2N, 2N, N)$ grid points for $(x,y,z)$ with $N$ being an even number. We employ the finite-volume scheme with a reflux prescription and Harten-Lax-van Leer contact (HLLC) Riemann solver, as that implemented in \cite{kiuc22}, for hydrodynamics evolution to better conserve the total baryon mass of the system.

For the outer boundary condition, we use the outgoing boundary condition for metric variables following \cite{shib95}
and specifically include an additional term for the scalar field variables $Q=(\varphi, \Phi)$ as
\begin{align}
    Q(t, r) = \left( 1 - \frac{\Delta r}{r} \right) Q(t-\Delta t, r - \Delta r) e^{- m_\phi \Delta r},
\end{align}
to capture the exponential decay tail due to the mass term $m_\phi$. 
Here, $\Delta r=c\Delta t$ with $\Delta t$ the time step in numerical computation. We test the convergence of our code in three different resolutions (see \cref{conv.test}). Unless otherwise specified, we adopt $N=94$ as the standard resolution of this paper which corresponds to $\Delta x=157$~m in the finest box.
The details of the numerical setup can be found in \cref{tab:numerical_setup} in \cref{conv.test}.

The primary purpose of this paper is to investigate how the scenarios of post-merger remnants depend on the binary mass, $B$, $m_\phi$, and the EOS while restricting ourselves to equal-mass binaries. However, rather than specifying the binary mass as the sum of the ADM masses of the two NS members, we identify the binary mass as the total \emph{rest mass},
\begin{align}
    M_b := \int \rho u^t \sqrt{- g} d^3 x,
\end{align}
contained in the binary. 
Taking into account the GW event GW170817, scalar masses of $m_\phi\gtrsim10^{-11}$~eV are favored unless the coupling constant $B$ is so small that the NSs in the observed system are non-scalarized \cite{kuan23}. This condition on $m_\phi$ is several orders of magnitude greater than the constraint concluded from the pulsar timing observations, which is $m_\phi\gtrsim10^{-15}$~eV \cite{rama16,yaza16}, while more rigorous Bayesian inference studies are required to transform the suggestion of $m_\phi\gtrsim10^{-11}$~eV into a constraint (for strong couplings). On the other hand, a mass of $m_\phi\agt2\times10^{-11}$~eV would significantly suppress scalarization in NSs since the associated Compton length is shorter than the stellar size. 
Aiming to study the scalar's influence on BNS mergers, we focus on cases where NSs can develop scalar cloud before and/or after merger, and thus the range of interest of $m_\phi$ is narrow. We will consider only one canonical value for the scalar mass, viz. $m_\phi=1.33\times10^{-11}$~eV ($\comp=14.8$~km), to quantitatively study how $B$ influences the lifetimes of the HMNSs ($\Tns$) and scalar cloud ($\Tsc$) in post-merger systems.

\begin{figure}
    \centering
    \includegraphics[width=\columnwidth]{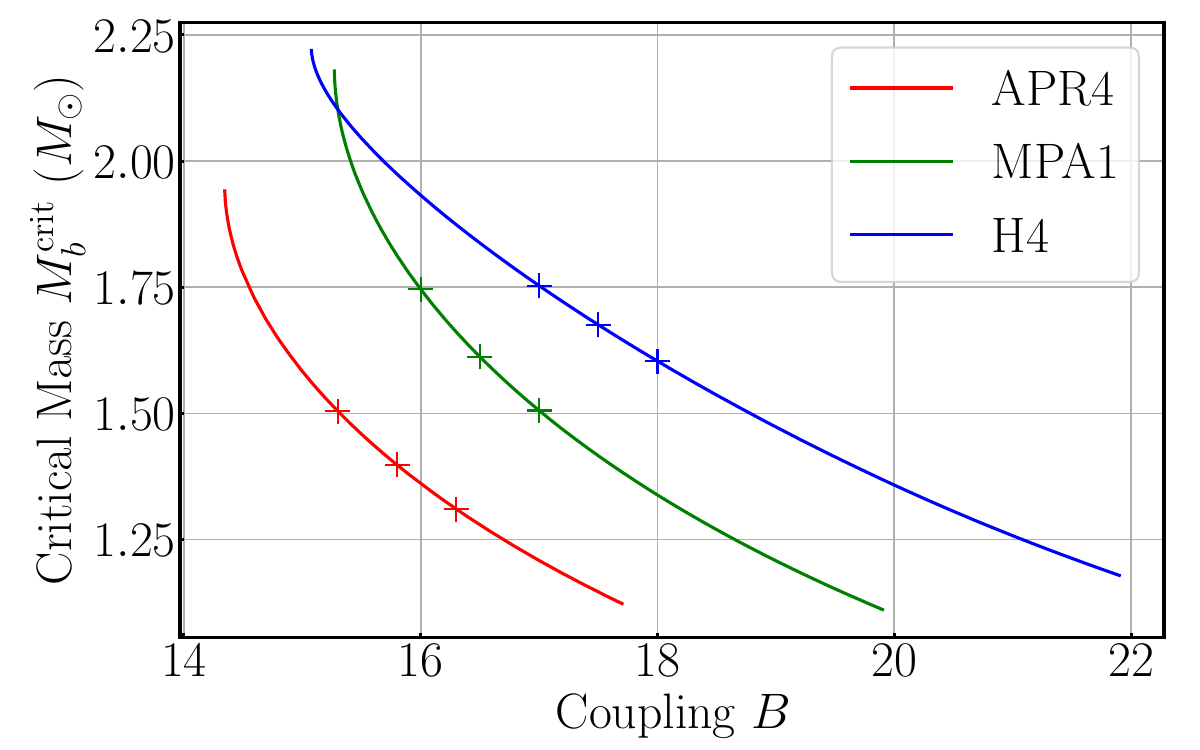}
    \caption{Critical baryon rest-mass of NSs that are marginally scalarized when isolated as functions of the coupling constant $B$ for  $m_\phi=1.33\times10^{-11}$~eV.
    The plus markers indicate the coupling strength which we choose to generate the mass sequences for each EOS.
    }
    \label{fig:bdry}
\end{figure}

For each EOS, we choose three different coupling strengthes $B$ such that an isolated NS with $M_b = 1.60 M_\odot$ would be either non-scalarized, marginally scalarized or spontaneously scalarized as illustrated in \cref{fig:bdry}.
We explore a wide range of NS's baryon mass spanning from $1.60 M_\odot$ to $1.90 M_\odot$ as summarized in \cref{modellist} with \cref{tab:outcomes_APR4,tab:outcomes_MPA1,tab:outcomes_H4} for APR4, MPA1 and H4 EOS, respectively, to investigate different outcomes of post-merger remnants. Each model is referred to in the manner of the example: \texttt{MPA1\_B16.5\_M1.70} corresponds to the equal mass binary with the MPA1 EOS, $B=16.5$, and $M_b = 1.70 M_\odot$ for an individual NS. Since the coupling strengthes considered are not very strong, the ADM mass ($M_{\rm ADM}$) of the isolated NS deviates only slightly ($\lesssim 10^{-3} ~M_{\odot}$) from the star having the same baryon mass in GR.

We construct the BNS initial data in a quasi-equalibrium state by generalizing the public spectral code FUKA \cite{Pape21} to the massive DEF theory. The BNS configurations are prepared with an initial separation of $44.31$~km, with which the BNS models experience 3--5 orbits before merger.
Note that in our numerical simulation, the virial error of the initial data defined by the relative difference of ADM mass and Komar mass, are always smaller than $0.04 \%$.
We refer the readers to Ref.~\cite{kuan23} for the detailed initial data formulation for constructing quasi-equilibrium states of BNS in the massive DEF theory.

\subsection{Gravitational Wave Extraction}

The information of GWs emitted is obtained by extracting the complex Weyl scalar $\Psi_4$ in the local wave zone (see, e.g., \cite{yama08,kiuc17,kiuc20} for details).
The Weyl scalar $\Psi_4$ is decomposed into $(l,m)$ modes with spin-weighted harmonics as
\begin{align}
    \Psi_4 (t_{\rm ret}) = \sum_{l,m} \Psi_4^{l,m} (t_{\rm ret}) {}_{-2} Y_{lm}(\theta, \phi),
\end{align}
where the retarded time $t_{\rm ret}$ is defined by \cite{hoto13d,kiuc17}
\begin{align}
    t_{\rm ret} := t - D - 2 M_{\rm inf} \ln \left( \frac{D}{2 M_{\rm inf}} - 1 \right).
\end{align}
Here, $M_{\rm inf} := M_{1, {\rm ADM}} + M_{2, {\rm ADM}}$ is the total ADM mass of the isolated NSs separated at spatial infinity
and $D$ is the areal radius of the extraction sphere approximated as \cite{kiuc17}
\begin{align}
    D \approx R_0 \left(1 + \frac{M_{\rm inf}}{2R_0} \right)^2,
\end{align}
by assuming isotropic coordinates of non-rotating black holes in the wave zone with $R_0$ being the corresponding coordinate radius.
We evaluate $\Psi_4$ at the finite radius $R_0=480~M_{\odot} \approx 709~{\rm km}$ and then analytically extrapolate the waveform toward null infinity by Nakano's method \cite{lous10,naka15a,naka15b}. We shall focus only on the dominant $(l,|m|)=(2,2)$ mode in this work because the contribution from other higher-multipole modes is minor for the equal-mass BNSs.
The harmonic mode of GWs can be evaluated by integrating $\Psi_4^{l,m}$ twice in time given by
\begin{align}
    h^{l,m}(t_{\rm ret}) &=  h_{+}^{l,m} (t_{\rm ret}) - i h_{\times}^{l,m} (t_{\rm ret}) \nonumber \\
    &= - \int^{t_{\rm ret}} dt' \int^{t'} \Psi_4^{l,m} (t'') dt'' \nonumber \\
    &= \int df' \frac{\tilde \Psi_4^{l,m}(f')}{ (2\pi \max(f', f_{\rm cut}))^2} e^{2 \pi i f' t_{\rm ret}},
\end{align}
where the last line shows the fixed frequency method of~\cite{reis11} we employed for the calculation and $f_{\rm cut}$ is the cutoff frequency set to be $0.8 M_{\rm inf} \Omega_0 / (2\pi)$ with $\Omega_0$ being the initial angular velocity of the binary obtained from the initial data.
The merger time $t_{\rm merge}$ is defined at the time of the peak GW strain $h^{2,2} := h_+^{2,2} -i h_{\times}^{2,2}$,
where $h_+^{2,2}$ and $h_{\times}^{2,2}$ are the plus and cross polarization of $l=m=2$ GWs, respectively. 
We also calculate the instantaneous frequency $f_{\rm GW}$ of the $(2,2)$ mode by
\begin{align}\label{eq:instan_fgw}
    f_{\rm GW} = \frac{1}{2\pi} {\rm Im}\left( \frac{h^{\ast 2,2} \dot{h}^{2,2}}{|h^{2,2}|^2}\right),
\end{align}
where the asterisk and dot symbols denote the complex conjugate and the time derivative, respectively.
The interval between $t_{\rm merge}$ and the apparent horizon formation time $t_{\rm AH}$ defines the lifetime of HMNSs (i.e., $\Tns:= t_{\rm AH} - t_{\rm merge}$), and the lifetime of the scalar cloud, $\Tsc$, is determined by the interval between the merger and the descalarization in the HMNSs (if at all).

We obtain the amplitude of the Fourier spectrum of GWs following \cite{kiuc09,hoto11}
\begin{align}
    \tilde h(f)^{2,2} = \sqrt{\frac{|\tilde h_+^{2,2}(f)|^2 + |\tilde h_\times^{2,2} (f)|^2}{2}},
\end{align}
from the Fourier transforms of plus $\tilde h_+^{2,2} (f)$ and cross $\tilde h_{\times}^{2,2}(f)$ polarization of GWs with $f$ being the GW's frequency.
The dimensionless effective amplitude $h_{\rm eff}(f)$ of GWs is defined by
\begin{align}
    h_{\rm eff}(f) := f \tilde h^{2,2}(f).
\end{align}

The propagation group velocity of scalar waves ($v_g$) is stretched by $m_\phi$, and the dispersion relation is given by \cite{kuan23,josu22}
\begin{align}
    v_g = \left( 1+ m_{\phi}^2 \lambdabar_{\rm gw}^2 \right)^{-1/2},
\end{align}
with $\lambdabar_{\rm gw}$ being the wavelength of the scalar wave.
For $\lambdabar_{\rm gw} \gg \lambdabar_{\rm comp}$, the speed of scalar waves is much lower than the speed of light, and thus, essentially prohibiting the emission of scalar waves~\cite{just12,roxa20}. In this work, we consider a zero asymptotic value for the scalar field ($\varphi_0 = 0$), and consequently, scalar waves do not couple to the interferometer leaving no extra mode such as the breathing and longitude modes in emitted GWs.

\section{Post-merger scenarios} \label{sec:postmerger}

\begin{figure}
    \centering
    \includegraphics[width=\columnwidth]{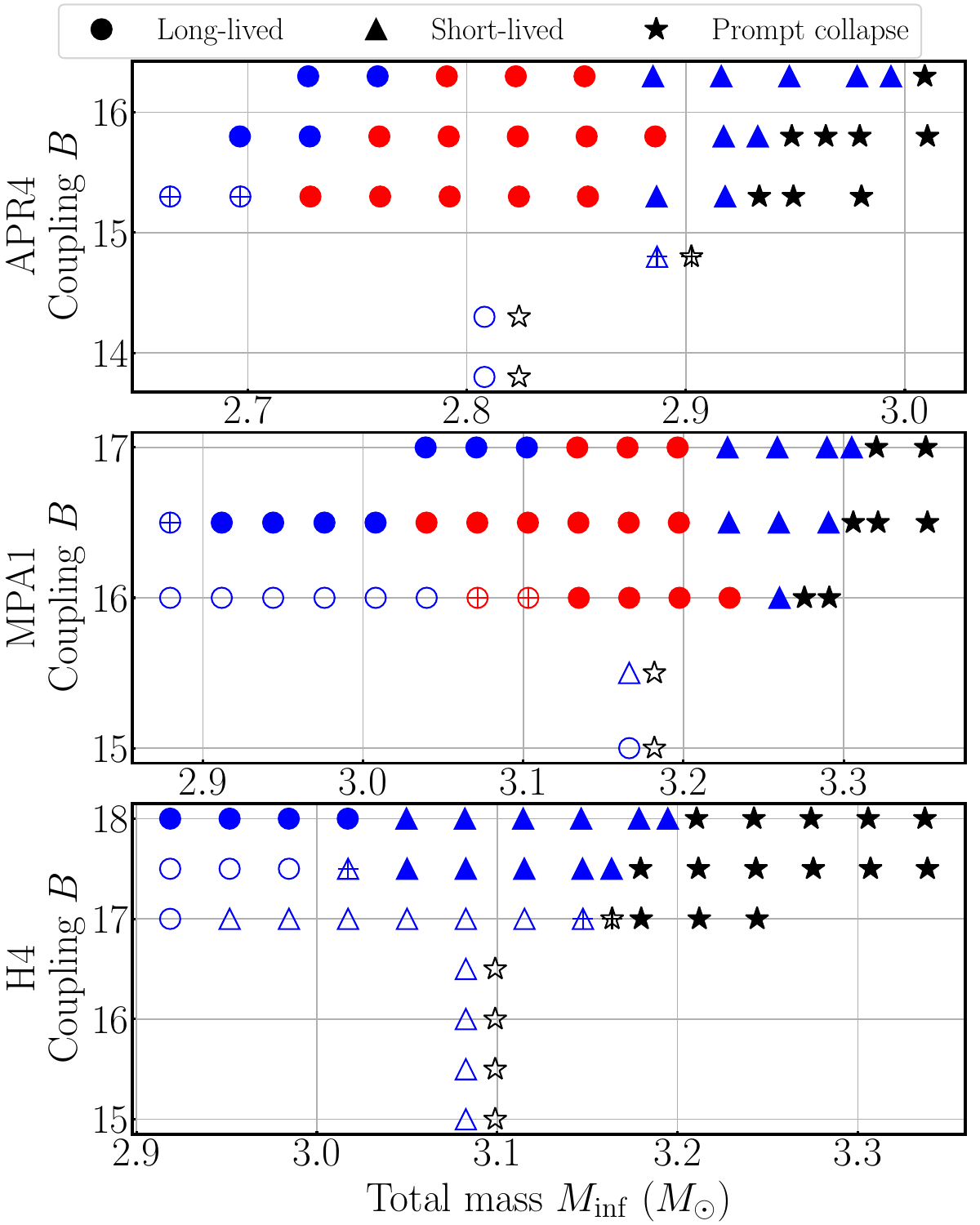}
    \caption{Summary of all the models in this work.
    The circle, triangle and black star markers represent the final fate of postmerger remnant as long-lived HMNSs, short-lived HMNSs, and prompt collapse to a black hole, respectively.
    The filled (resp. hollow) markers indicate the presence (resp. absence) of spontaneous scalarization for isolated NS while the plus markers indicate the occurrence of dynamical scalarization. The models that undergo descalarization are marked in the red color.
    }
    \label{fig:summary}
\end{figure}

In GR, the final fate of the post-merger remnant of BNSs depends primarily on the total mass and the EOS, while the mass of dynamical ejecta and the torus formed around the post-merger black hole (if at all) should be also sensitive to the mass ratio \cite{shib03,rezz10,diet17}. 
In terms of the HMNS's lifetime, we categorise the final outcome of BNS mergers into three different scenarios:
\begin{enumerate}[(i)]
    \item prompt collapse to black hole,
    \item short-lived HMNS formation $(\Tns < 10~{\rm ms})$,
    \item long-lived HMNS formation $(\Tns > 10~{\rm ms})$,
\end{enumerate}
where the criteria of $10$~ms is a subjective choice. On top of the above categorization for BNS remnants, the presence of a scalar field introduces more variety in the final states (see \cref{fig:summary}).

\begin{figure}
    \centering
    \includegraphics[width=\columnwidth]{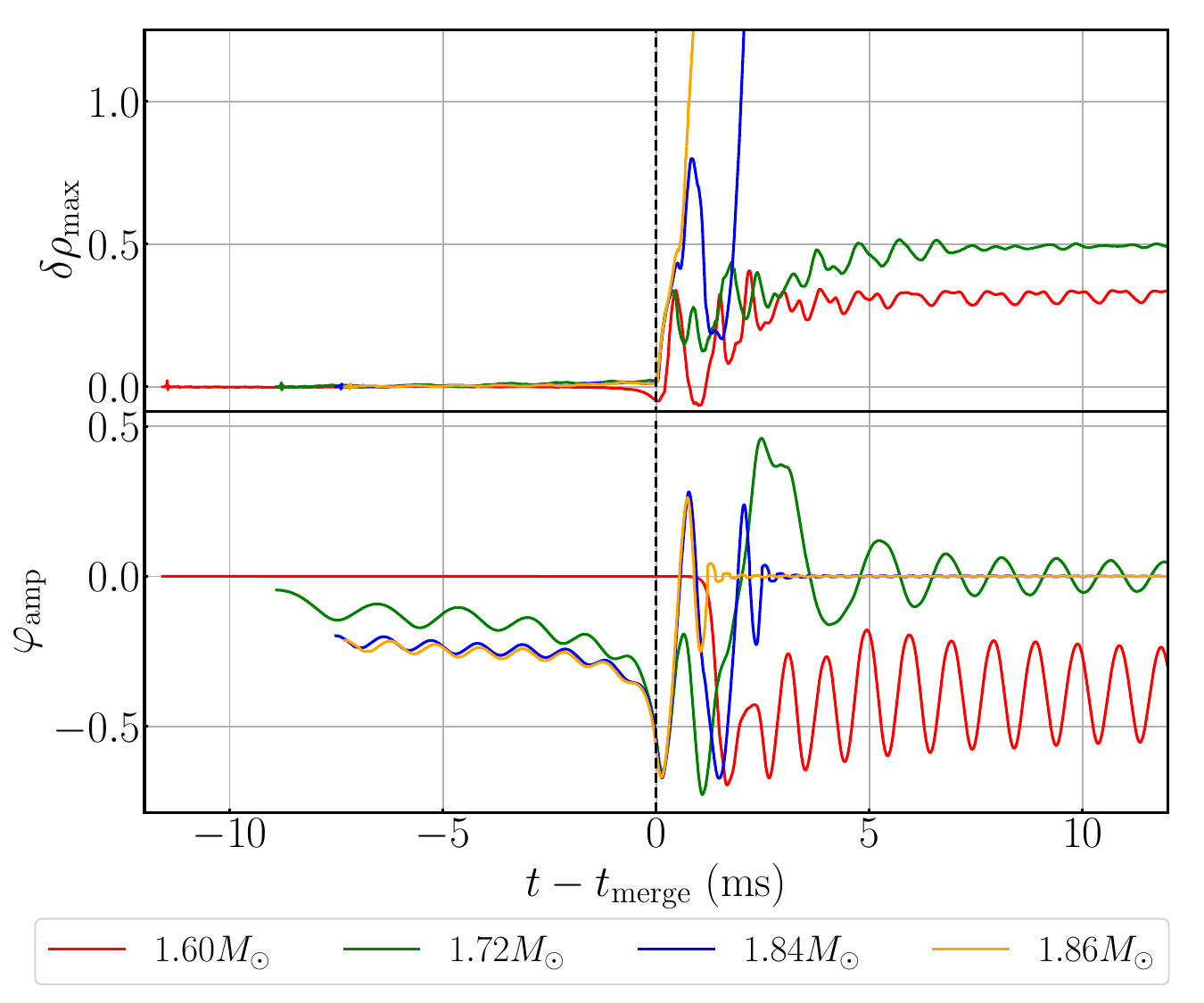}
    \caption{Evolution of the relative difference of maximum rest-mass density $\delta \rho_{\rm max}:= \rho_{\rm max}(t)/\rho_{\rm max}(t=0) - 1$ (top)
        and of the maximum scalar field amplitude $\varphi_{\rm amp}$ (bottom) with different initial baryon rest mass of individual NS with MPA1 EOS. The scalar-field parameters are set as $B=16$ and $m_\phi=1.33\times10^{-11}$~eV.
    }
    \label{fig:rhot}
\end{figure}

All the possible outcomes are showcased in \cref{fig:rhot}, where the evolution of the relative difference of maximum rest-mass density,
\begin{align}
    \delta \rho_{\rm max} := \rho_{\rm max}(t)/\rho_{\rm max}(t=0) - 1,
\end{align}
and the maximum scalar-field amplitude\footnote{Since the change of sign of $\varphi \rightarrow - \varphi$ does not alter the evolution of the system, we adopt the convention of negative value of $\varphi$ when spontaneous scalarization happens. Therefore, we flip the sign of $\varphi$ in the plots if positive $\varphi$ arises when the HMNS is spontaneously scalarized unless $\varphi$ experiences change of sign in the scalarization/descalarization process.}, 
\begin{align}
    \varphi_{\rm amp} := {\rm sgn}(\varphi) \max(|\varphi|),
\end{align}
are plotted for four selected models with MPA1 EOS and scalar-field parameters $(B, m_\phi) = (16, 1.33\times 10^{-11}{\rm ~eV})$.
\begin{figure*}
    \centering
    \includegraphics[width=\textwidth]{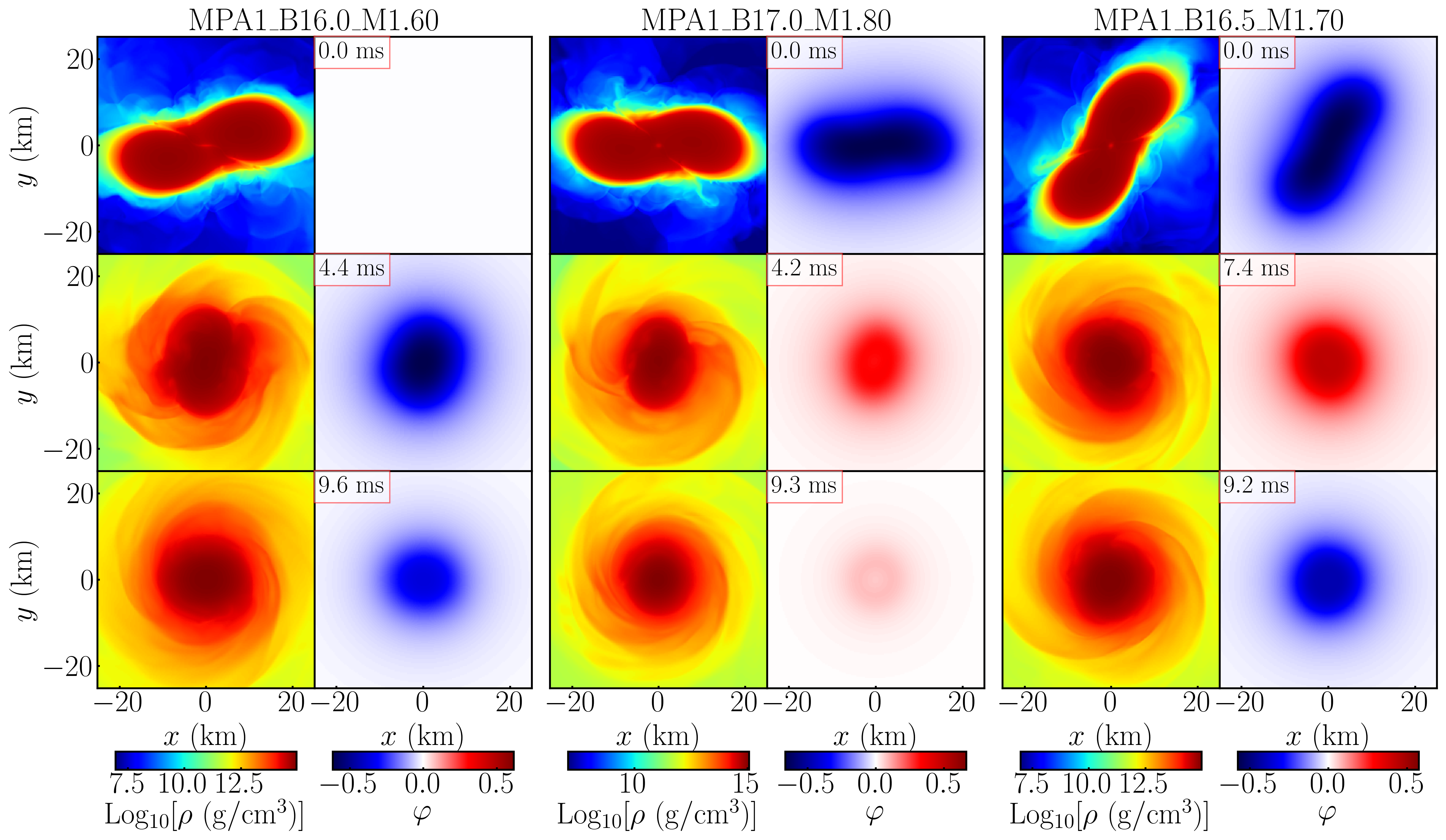}
    \caption{Snapshots of the rest-mass density $\rho$ (left column of each panel) and scalar field $\varphi$ (right column of each panel) on the equatorial plane for the cases of long-lived HMNS formation with the MPA1 EOS.
    The baryon mass $M_b$ of each NS in units of $M_\odot$ and coupling strength $B$ are 
    $(M_b, B) = (1.60, 16.0)$ (left)
    $(M_b, B) = (1.80, 17.0)$ (middle)
    $(M_b, B) = (1.70, 16.5)$ (right).
    The time for each snapshot is indicated in the red boxes with time measured from the onset of merger.
    }
    \label{fig:long_lived_HMNS}
\end{figure*}
We briefly summarize all the possible scenarios of the scalar field evolution according to \cref{fig:rhot}, and leave the in-depth discussion to the following sections. In the pre-merger phase, the scalar field can be excited if the NSs are compact enough to undergo spontaneous scalarization (blue and yellow lines) or dynamical scalarization (green line). Otherwise, the scalar field remains  insignificant up to merger (red line).
As we will show in \cref{sec:threshold_mass}, the scalarization history of the BNS plays an important role in the prompt-collapse threshold mass.
In the post-merger phase, depending on the final mass of the HMNS, it can either be spontaneously scalarized (red) or "descalarize" after a certain time to form an oscillating scalar cloud with appreciable amplitude. In the case where black holes are formed (blue and yellow), the scalar field does not dissipate away entirely, and an oscillating scalar cloud forms from the fossil scalar field instead. 
Although we will discuss different outcomes of BNS mergers based on the lifetimes of the HMNS and the scalar cloud, it should be noted that these timescales are not to be taken as exact for simulated models. In fact, it is impossible to determine accurately the lifetimes in the numerical simulation in practice since the HMNS after the merger is close to a marginally stable state, and any small perturbation (including numerical errors) will alter its collapse time and thus the dynamics is extremely sensitive to the grid resolution. Thus, the values can be considered as an approximate estimate and the scenarios characterized by them are still qualitatively robust.

It can be noticed that the scalar field $\varphi_{\rm amp}$ experiences $\sim 10\%$ perturbation for scalarized binaries in the inspiral phase, which
indicates that the scalar field has not yet perfectly reached the quasi-equilibrium state.
One possible reason is the insufficient grid resolution to resolve the exponential falloff tail of the scalar field in our initial data solver. The other possible reason is that the zero scalar field "momentum" $\Phi = 0$ condition employed in our initial data formulation \cite{kuan23} could possibly induce some initial perturbation in the system.
While any initial perturbation of the scalar field in the massless DEF theory~\cite{tani15,shib14} can freely propagate out and dissipate quickly,
in the presence of non-zero scalar mass $m_\phi$ perturbations with a wavelength smaller than the Compton wavelength will be trapped and remain in the vicinity of the system. Nonetheless, the initial perturbation of the rest-mass density $\delta \rho_{\rm max}$ is less than $1 \%$, and hence, we believe that the effect of the scalar field perturbation is minor.

\subsection{Long-lived neutron star remnant}
\label{sec:Long-lived_HMNS}

We first recap the key criterion for spontaneous scalarization in a single star following \cite{kuro23,shib14,damo93}, which is also useful in explaining the evolution of the scalar field in the HMNS.
The onset of scalarization can be approximately described by taking the weak field limit of \cref{eq:s_eq} with an average value of $T$ within the star radius $R$, $\bar T$, as
\begin{align} \label{eq:scalarization}
    (\Delta - m_\phi^2) \varphi = 2 \pi B \bar T \varphi,
\end{align}
where $\Delta$ is the flat Laplacian. 
%Depending on the coupling strength $B$, $T=-\rho h + 4P$, and the compactness of the star,
Denoting $k^2:= - (2\pi B \bar T + m_\phi^2)$, the conditions for scalarization are given as $k^2 > 0$ and $k R \rightarrow \pi / 2$ for $R$ the NS's radius \cite{shib14,kuro23}.
%while for $k^2 < 0$ scalarization is not likely to happen.
For the case of $B>0$ and assuming that the relativistic corrections to matter are small (i.e., $\bar T\sim - \rho$), scalarization is likely to happen if $\bar T\sim - \rho < T_{\rm crit}:=- m_\phi^2 / (2 \pi B)$. 
However, scalarization is unlikely to occur if a bulk of HMNS's interior is  ultrarelativistic with $T=-\rho h +4P > T_{\rm crit}$. 
The critical value of $\bar T$ depends on the actual profile of the star, while $T_{\rm crit}$ still serves as a good indicator for understanding the scalarization criterion (see below).

Shortly after the merger, an ultrarelativistic region can be formed in the HMNS for some cases, where the descalarization soon ensues. However, the core of a natal HMNS may not be in an ultrarelativistic regime even though possessing a much higher central density than that of the progenitors. In this case, scalarization may occur in the HMNS even if the progenitors remain unscalarized up to the merger (i.e., for a not-extremely large value of $B$). However, the subsequent mass accretion may lead to the emergence of a region with $\bar T>T_{\rm crit}$, resulting in a descalarization.
In the event of a \emph{marginal} descalarization, the scalar cloud trapped by the central object oscillates with a larger amplitude than the case where the condition of $\bar T>T_{\rm crit}$ is conspicuously satisfied. 

Before delineating different scalarization and descalarization scenarios for long-lived HMNSs in the following subsections, we demonstrate each channel by a representative model in \cref{fig:long_lived_HMNS}, in which the snapshots of rest-mass density (left column of each panel) and scalar field (right column of each panel) on the equatorial plane in the post-merger phase are displayed. 
For \texttt{MPA1\_B16.0\_M1.60}, the HNMS never reaches the ultrarelativistic regime and remains scalarized until the end of the simulation, while the descalarization upon the criterion is met fully and marginally for \texttt{MPA1\_B17.0\_M1.80} and \texttt{MPA1\_B16.5\_M1.70}, respectively.

\begin{figure}
    \centering
    \includegraphics[width=\columnwidth]{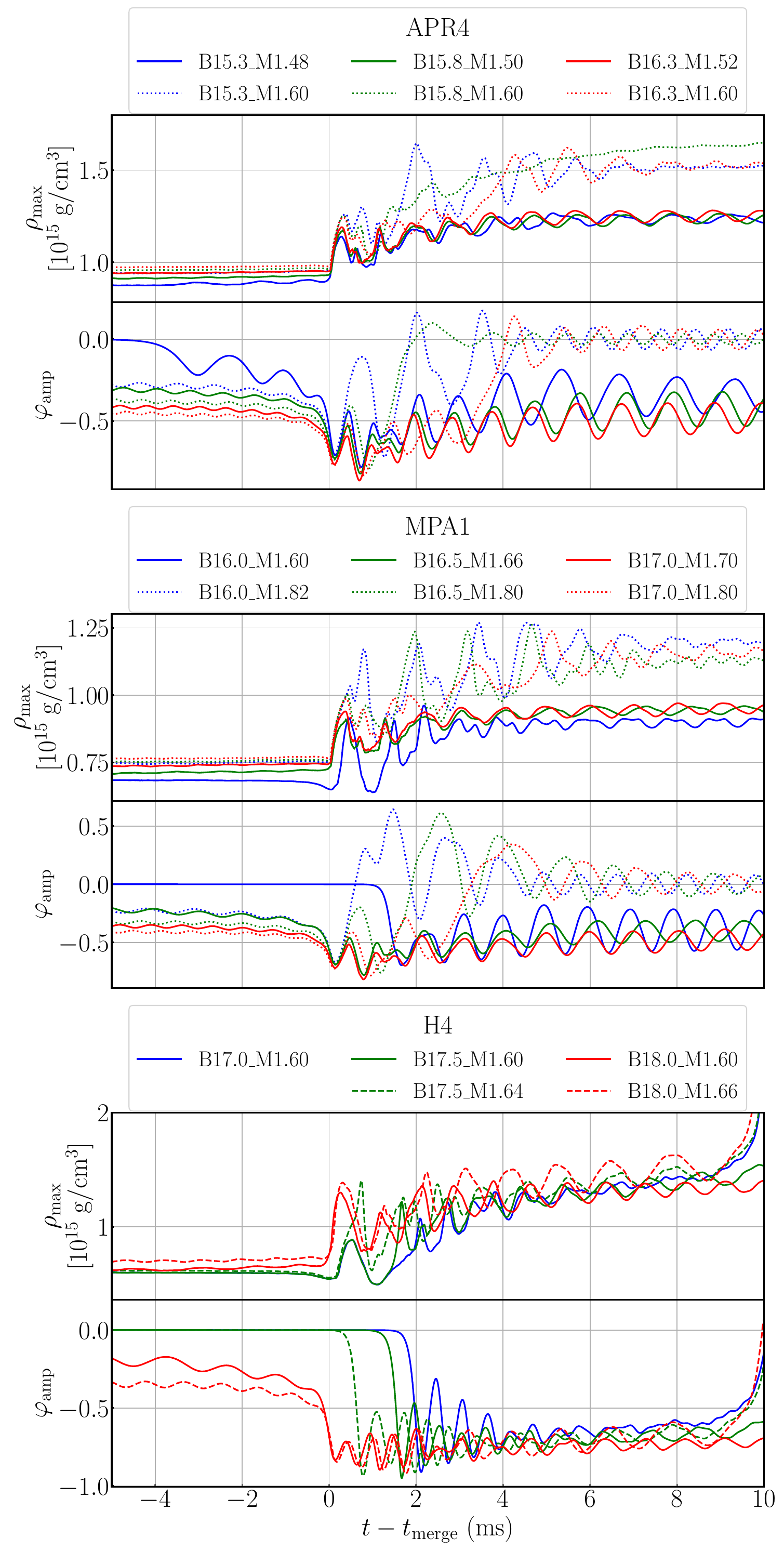}
    \caption{Evolution of the maximum rest-mass density $\rho_{\rm max}$ 
    and scalar-field amplitude $\varphi_{\rm amp}$ for the long-lived HMNS formation with APR4 (top), MPA1 (middle), and H4 (bottom) EOS.
    The solid and dashed curves correspond to spontaneously scalarized HMNS formation,
    and the dotted curves correspond to models in which descalarization happens within 10\,ms after the merger.
    }
    \label{fig:longlive_evo}
\end{figure}

\subsubsection{Long-lived scalarized HMNS}

\cref{fig:longlive_evo} shows the evolution of the maximum rest-mass density $\rho_{\rm max}$ and scalar-field amplitude $\varphi_{\rm amp}$ for selected models that yield a long-lived HMNS for three different EOSs. 
We first focus on the cases for which the HMNS confidently (solid) and marginally (dashed; present only for the H4 EOS) remains spontaneous scalarized at $t-t_{\rm merge}=10~{\rm ms}$.
Some models with small values of $B$ do not exhibit dynamical scalarization during the inspiral phase, but scalarization can still occur in the post-merger phase (e.g., \texttt{MPA1\_B16.0\_M1.60}), because the HMNS has a higher compactness compared to the corresponding isolated NS so that even for a small value of $B$, $\sqrt{-T}R \sim \sqrt{\rho} R \sim \sqrt{M/R}$ in the resulting HMNS can be high enough to fulfill the criterion of spontaneous scalarization.

Generally, the scalar field for scalarized HMNSs first gets amplified during merger and then settles down to a certain saturation level ($|\varphi| \sim 0.5-0.7$ in our cases) in a time interval of $\sim2$~ms. The exact timescale depends on the coupling strength $B$; for example,
the scalar field for \texttt{H4\_B17.5\_M1.60} takes $\approx 1.7$~ms to grow to the peak value after merger, while for \texttt{H4\_B17.0\_M1.60} it takes $\approx 2.1$~ms.
This illustrates that it typically takes longer for the scalar field to grow to saturation for a weaker coupling, in line with the previous numerical studies where massless scalar field is considered \cite{shib14}. 

The enhancement or activation of the scalar field during merger introduces an oscillation for it in the HMNS. Due to the non-zero mass of the scalar field, this oscillation does not dissipate quickly in contrast to the massless case~\cite{shib14},
but instead gets trapped and persists for a timescale longer than 10\,ms after the onset of merger with appreciable oscillation amplitude $\lesssim 0.1$ for $\varphi$.
The oscillation frequency of the scalar field coincides with the one for the rest-mass density at around 1~kHz. The mode associated with this pattern is believed to attribute to the radial $\phi-$mode since it falls in the band of a radial mode \cite{mend18} of scalarized HMNSs.

For \texttt{H4\_B17.0\_M1.60} (blue solid line), \texttt{H4\_B17.5\_M1.64} (green dashed), and \texttt{H4\_B18.0\_M1.66} (red dashed) in the bottom panel of \cref{fig:longlive_evo}, we find a unique feature. 
For these models, the scalar fields go to zero at $\sim 10$\,ms after the onset of merger, and a black hole forms very soon afterward as we can see that the rest-mass density is also growing rapidly. The descalarization shortly prior to the black hole formation is not triggered by the criterion $\bar T > T_{\rm crit}$, but rather should be attributed to the no-hair theorem in the DEF theory (e.g., \cite{soti15} and the references therein).

\subsubsection{Descalarized HMNS}

In this section, we pay attention to the models for which the long-lived HMNSs undergo descalarization that is induced by the secular contraction of the HMNS due to the GW emission and angular momentum redistribution via gravitational torque associated with the non-axisymmetric structure of the merger remnant.

The dotted curves in \cref{fig:longlive_evo} show the evolution of models that descalarize over a dynamical timescale after the onset of merger.
Taking \texttt{MPA1\_B16.0\_M1.82} as an example (blue dotted curve in the middle panel), we find that the scalar field promptly goes to zero when the maximum density rises to become ultrarelativistic during the post-merger evolution.
However, the scalar field does not stop at zero but instead form an oscillating scalar cloud around the HMNS with an appreciable amplitude of $\lesssim 0.1$,
which differs from the massless case in which the scalar field is completely turned off after descalarization~\cite{shib14}.
A note is necessary here to say that the term ``descalarized HMNS'' does not mean the scalar field is totally dissipated, but rather, it represents an HMNS with a long-lived oscillating scalar field with the zero time-averaged value $\left < \varphi_{\rm amp} \right >=0$.
Owing to the residue scalar cloud, it is non-trivial to determine definitely the time when descalarization happens, and we simply define the descalarization time $\Tsc$ as the time of the first zero crossing of the scalar field amplitude $\varphi_{\rm amp}$ during the post-merger phase.

\begin{figure}
    \centering
    \includegraphics[width=\columnwidth]{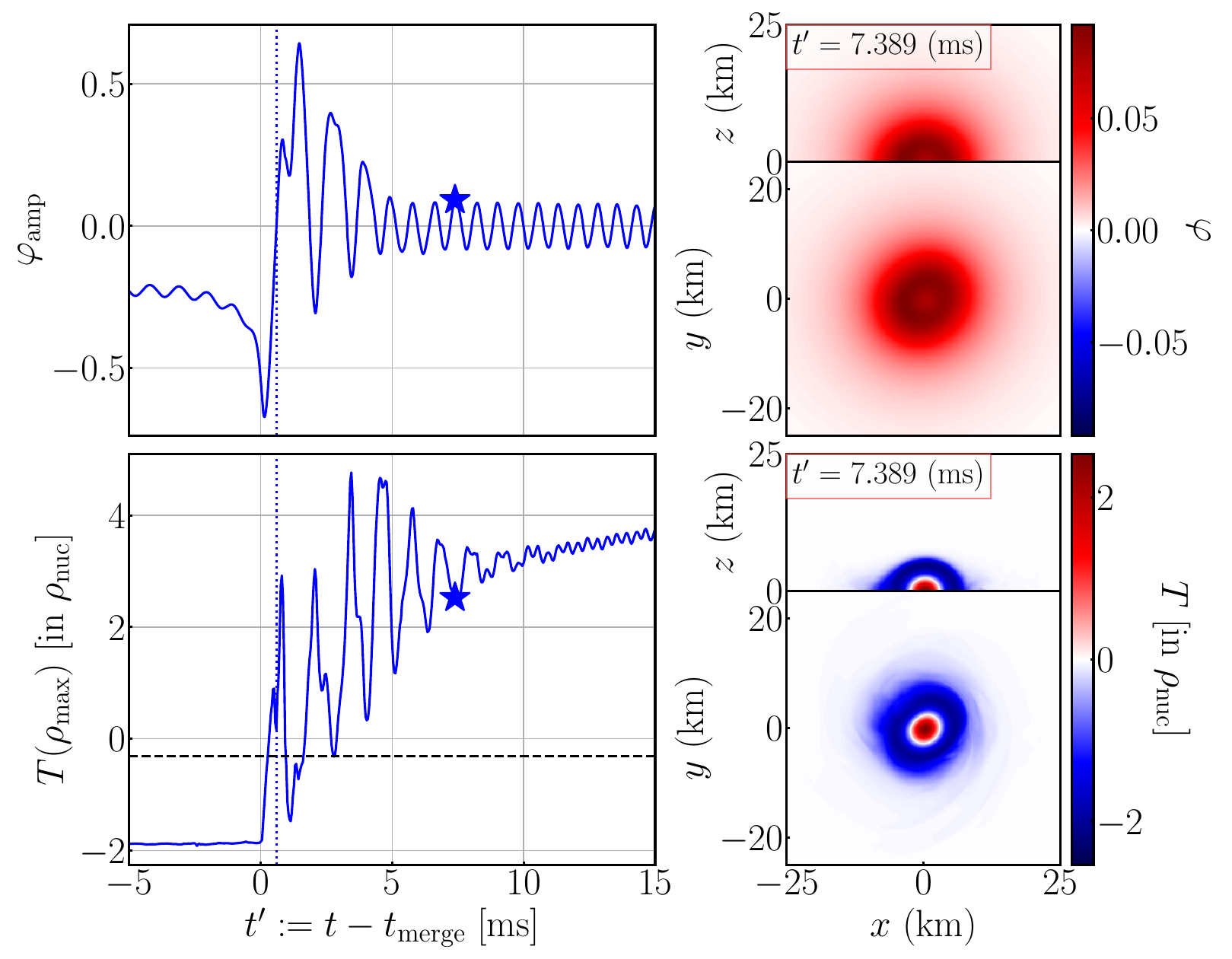}
    \caption{For a prompt descalarization scenario \texttt{MPA1\_B16.0\_M1.82}, left panels show the evolution of the scalar-field amplitude $\varphi_{\rm amp}$ (top left) 
    and trace of stress-energy momentum tensor $T:=T^a{}_a$ (bottom left) in units of the nuclear saturation density $\rho_{\rm nuc}= 2 \times 10^{14} ~{\rm g}/{\rm cm}^3$ at maximum density $\rho_{\rm max}$ point.
    The blue dotted vertical line and the black dashed horizontal line show the descalarization time $\Tsc$ and the critical value $T_{\rm crit}$, respectively.
    The blue stars indicate the time of the snapshots of $\varphi$ (top right) and $T$ (bottom right) on equatorial ($x$-$y$) and vertical ($x$-$z$) planes.
}
    \label{fig:traceT}
\end{figure}

To further understand the condition of the descalarization, we show the evolution of the scalar-field amplitude $\varphi_{\rm amp}$ together with $T$ at the maximum density, $T(\rho_{\rm max})$, in \cref{fig:traceT} for model \texttt{MPA1\_B16.0\_M1.82} for which the prompt descalarization happens during the post-merger phase with $\Tsc = 0.60 ~{\rm ms}$.
Here, $T(\rho_{\rm max})$ in units of the nuclear saturation density $\rho_{\rm nuc}$ ($=2\times 10^{14}\,{\rm g}/{\rm cm}^3$) is plotted. %with $c=1$.
In the inspiral phase, the NSs are initially spontaneously scalarized which is consistent with the scalarization condition $T(\rho_{\rm max})<T_{\rm crit}$ as the central value of $T \approx -2 \rho_{\rm nuc}$.
Once the NSs merge, $T(\rho_{\rm max})$ raises rapidly due to the increase in maximum density and thermal contribution from shock heating, and immediately flips sign to become positive. Soon after the scalar field crosses $T_{\rm crit}$, the descalarization occurs.
Note that $T(\rho_{\rm max})$ fluctuates around $T_{\rm crit}$ for a few times due to the radial oscillation, temporally satisfying the scalarization condition ($T<T_{\rm crit}$) during those cycles. As it turns out, the scalar field is likely to temporarily reach a high value as in spontaneous scalarization, and hence introduces large oscillation after $\Tsc$.

As $T(\rho_{\rm max})$ shifts further away from $T_{\rm crit}$, the scalar field quickly damps, leaving an oscillating scalar cloud around the HMNS. In contrast to the $\phi$-mode in spherical NSs in the massless DEF theory, for which the damping time of $\varphi$ is $\lesssim 1$~ms \cite{mend18}, the residual scalar cloud persists for more than 10~ms in the massive case, forming a long-lived quasi-normal mode with appreciable amplitude $\sim \mathcal{O}(0.1)$. Such a long-lived $\phi$-mode observed in both scalarized and descalarized cases is consistent with the results of \cite{blaz20}, which suggests that the presence of mass term $m_\phi$ could significantly extend the lifetime of the radial $\phi-$mode in the massive Brans-Dicke scalar-tensor theory.
Also shown in the right panels of \cref{fig:traceT} are the snapshots of the scalar field $\varphi$ and $T$ at 7.389~ms after the onset of merger. Despite of the large value of $T\sim 2 \rho_{\rm nuc}$ at the center which forbids the HMNS from being spontaneously scalarized, it still contains considerable matter with $T<T_{\rm crit}$ surrounding the center, whose size is comparable to the Compton wavelength $\lambdabar_{\rm comp}=14.8$~km. This creates an off-centered potential well in the right-hand-side of \cref{eq:scalarization} and as such traps the scalar field in a hollow sphere shape as shown in \cref{fig:traceT}, which is different from the scalar field profile of spontaneous scalarized HMNSs in \cref{fig:long_lived_HMNS}, for which the peak value of $\varphi$ is located at the center of the NSs (see also \cite{stay23}).

\begin{figure}
    \centering
    \includegraphics[width=\columnwidth]{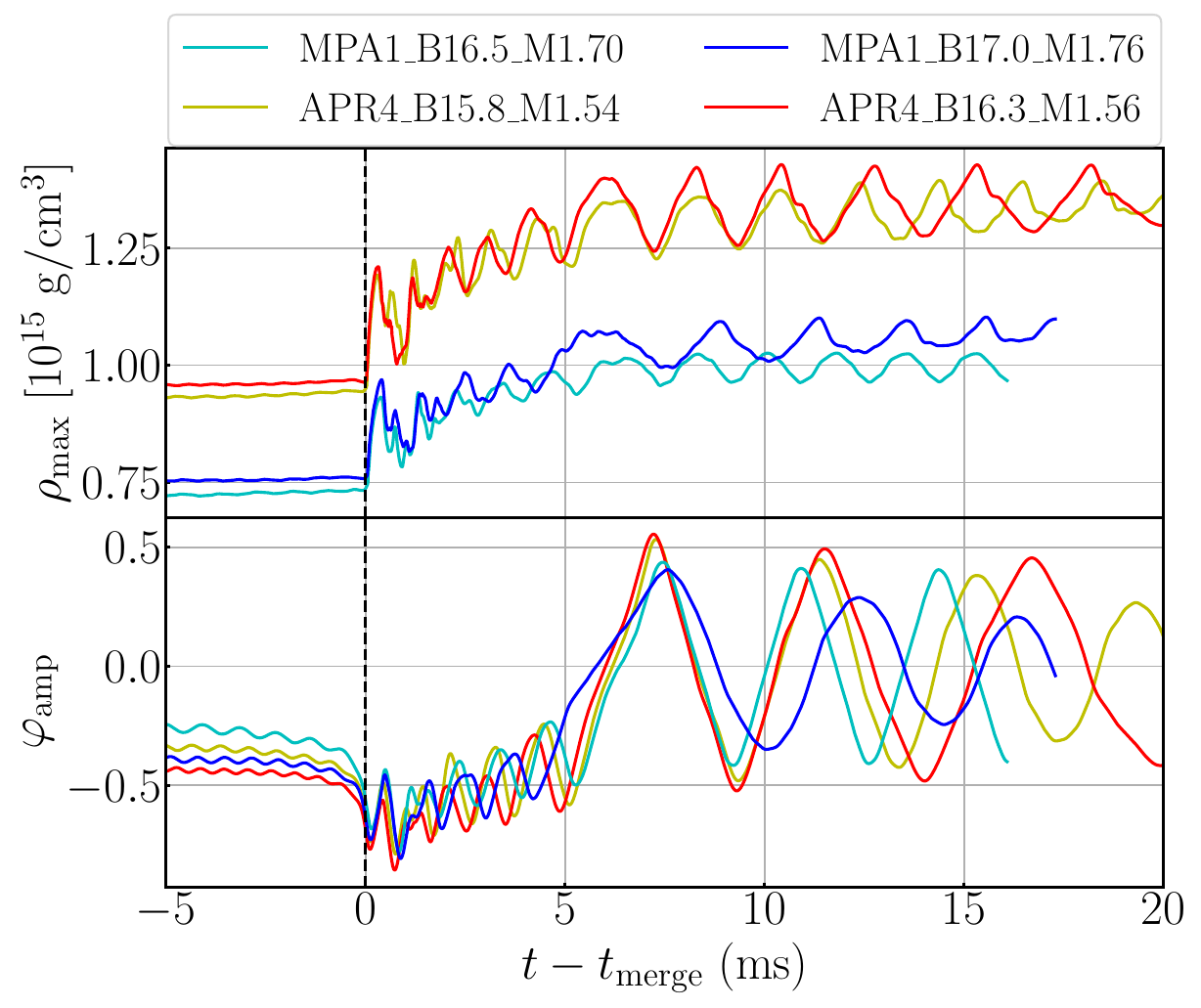}
    \caption{Evolution of the maximum density $\rho_{\rm max}$ (top) and scalar-field amplitude $\varphi_{\rm amp}$ for marginally descalarized models with APR4 and MPA1 EOSs.
    }
    \label{fig:marginal_evo}
\end{figure}

Other than the prompt descalarization scenario, the HMNS can still be subsequently descalarized due to the secular contraction. %\addms{$\leftarrow$Again do not write anything by prejudice. I do not think that accretion is the reason.}
In some models shown as red dotted curves in \cref{fig:longlive_evo}, such as \texttt{APR4\_B16.3\_M1.60} and \texttt{MPA1\_B17.0\_M1.80}, the HMNSs remain spontaneously scalarized for a few ms after the onset of merger. Meanwhile the rest-mass density $\rho_{\rm max}$ continues increasing due to the contraction resulting from the angular momentum dissipation by the GW emission and the angular momentum redistribution via gravitational torque associated with the non-axisymmetric structure of the HMNS until it reaches the ultrarelativistic limit and triggers the descalarization. However, if the maximum rest-mass density of the HMNS settles down to a value very close to the critical value for scalarization,
the HMNS may undergo several cycles going between states of scalarization and descalarization due to the density fluctuation caused by the radial oscillation.
\cref{fig:marginal_evo} shows the evolution of maximum density $\rho_{\rm max}$ and scalar-field amplitude $\varphi_{\rm amp}$ for the marginally descalarized models, which are denoted as the least massive descalarized HMNS along the mass sequence. As the transition state between scalarized and descalarized HMNSs, any perturbation in density allow the HMNS to temporarily reach the scalarization criteria and drive the scalar field towards the level of the spontaneously scalarized HMNS. Different from the hollow spherical scalar clouds formed around the descalarized HMNS, the scalar cloud's profile still peaks at the center, similar to the spontaneously scalarized models in the marginally descalarized as illustrated in \cref{fig:long_lived_HMNS} for model \texttt{MPA1\_B16.5\_M1.70} (right panel). Therefore, it contains a much stronger oscillation in $\varphi$ than for other descalarized models with the amplitude $\sim 0.5$.

In addition to the strong scalar-cloud oscillation, the marginally descalarized models also have a much lower frequency of $\phi$--mode with $\lesssim 500$~Hz. We perform Fourier transform of $\varphi_{\rm amp}^2$ for the post-merger phase of long-lived HMNSs to obtain the characteristic frequency\footnote{
Note that instead of the conventional choice $\varphi_{\rm amp}$ used in other studies \cite{mend18,blaz20}, we choose specifically $\varphi_{\rm amp}^2$ for the Fourier analysis which introduces an extra factor of 2 in frequency for the perturbation of $\varphi$ if the background scalar field is zero (i.e. in the case of a descalarized HMNS with time-averaged $\left< \varphi_{\rm amp}\right>=0$). However, this choice does not alter the frequency of the Fourier spectrum for the spontaneously scalarized HMNS case.}
since the scalar field enters the modified Einstein field equations,  \cref{eq:einstein}, as $\phi \sim \varphi^2$ and thus $\varphi^2$ is physically more relevant to hydrodynamics.
Indeed, we find a better agreement between the Fourier spectrum of $\rho_{\rm max}$\footnote{While the perturbation of rest-mass density $\rho$ is decoupled with $\varphi$ in the GR branch of static spherical stars \cite{mend18},
the evolution of $\rho$ would still be affected by $\varphi$ even for the descalarized HMNS case in full dynamical simulation.} and $\varphi_{\rm amp}^2$.
To obtain a cleaner spectrum, we cut the transient evolution of scalar field after the change of the scalarization state, which is the first 2~ms after the onset of merger for the scalarized cases, while for the descalarized cases we cut the first few ms after the descalarization happened until the scalar field reaches at most twice of its final amplitude.

\begin{figure}
    \centering
    \includegraphics[width=\columnwidth]{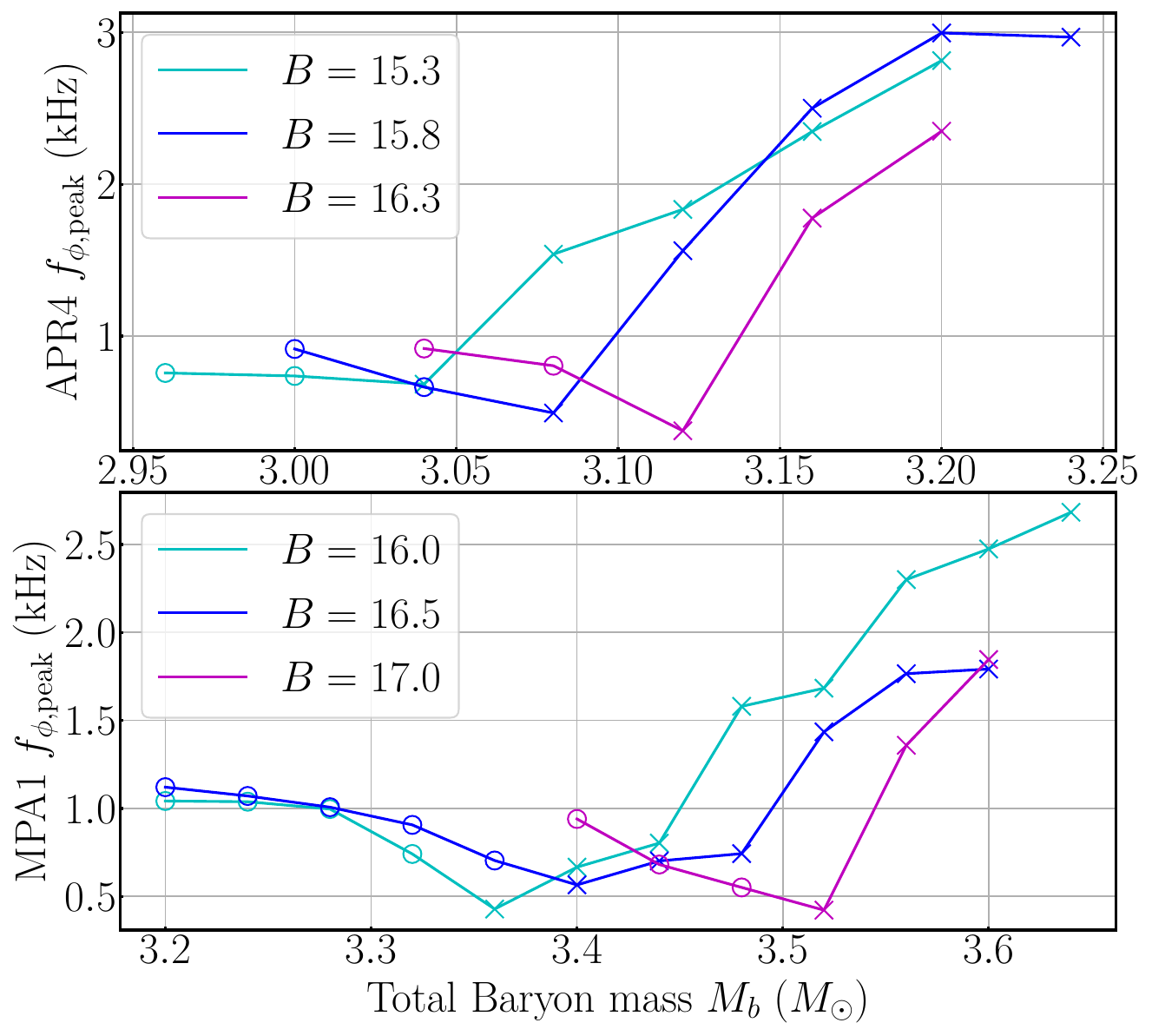}
    \caption{The peak frequency of $\varphi_{\rm amp}$ for EOS APR4 (top) and MPA1 (bottom) with respect to the total baryon mass of the system for the long-lived HMNS scenario. The cross and circle markers indicate the models with and without descalarization, respectively.
    }
    \label{fig:phimode}
\end{figure}

Denoting $f_{\phi, {\rm peak}}$ as the peak frequency of the Fourier spectrum of $\varphi_{\rm amp}^2$, which is believed to be the $\phi$-mode of the HMNS, \cref{fig:phimode} summerizes how $f_{\phi, {\rm peak}}$ varies along the mass sequence for the APR4 and MPA1 EOSs, for which a descalarized HMNS can be formed. The cross and circle markers indicate the models with and without descalarization,  respectively. As the total baryon rest-mass of the scalarized HMNS increases, $f_{\phi, {\rm peak}}$ drops and eventually reaches its minimum at the marginally descalarized models. After that, $f_{\phi, {\rm peak}}$ rises along the mass sequence for the descalarized HMNS. This is consistent with the characteristics of $\phi$-mode as shown in Fig.~2c in \cite{mend18} for which the mode frequency of the spontaneously scalarized branch first drops to zero at the bifurcation point, indicating the end of the scalarized state due to the mode instability, and then rises again in the GR branch. Therefore, we believe that the dominant mode in $\varphi_{\rm amp}^2$ is the radial $\phi$-mode and the zero-frequency point of $f_{\phi, {\rm peak}}$ at the marginally descalarized model indicates the bifurcation point of scalarized and GR branches.

\begin{figure}
    \centering
    \includegraphics[width=\columnwidth]{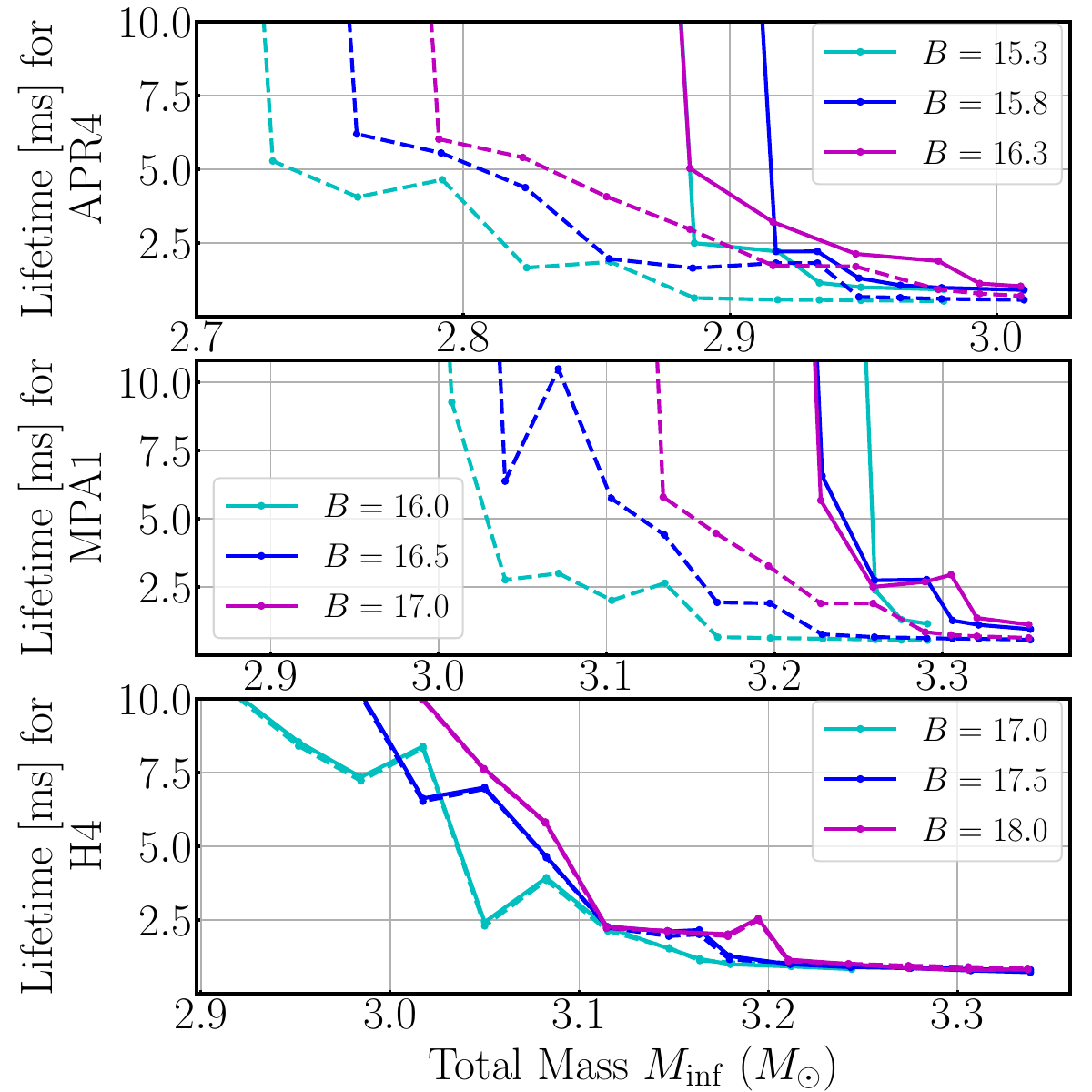}
    \caption{Lifetimes of the excited scalar field (dashed) and HMNS (solid) for the DEF theory with $m_\phi=1.33\times10^{-11}$~eV while various coupling strengths as functions of the total mass $M_{\rm inf}:=M_{\rm ADM, 1} + M_{\rm ADM, 2}$ of NSs for APR4 (top) MPA1 (middle), and H4 (bottom) EOSs.
    }
    \label{fig:lifetime}
\end{figure}

We summerize $\Tns$ (solid) and $\Tsc$ (dashed) for the simulated models whenever they can be determined in \cref{fig:lifetime}.
The scalar cloud's lifetime $\Tsc$ depends strongly on the coupling strength $B$ as shown by the dashed curves in \cref{fig:lifetime}.
In general, $\Tsc$ is longer for the larger values of $B$.
It is noticed that the descalarization of HMNSs only occurs in APR4 and MPA1 EOSs, while all the models with the H4 EOS (bottom panel) only descalarize right before the collapse, i.e., the lifetimes $\Tns$ and $\Tsc$ overlapped with each other. Although we pick up weak coupling strengths that induces the scalarization for the static spherical NSs, ranging from $B=17$ to 18 for H4 EOS, the critical coupling strength $B$ for the marginally scalarization decreases rapidly for more massive NSs as shown in \cref{fig:bdry}.
For static spherical NSs with total baryon mass greater than $2 M_{\odot}$, spontaneously scalarization can happen for much lower coupling $B < 16$ in H4 EOS, and we expect such critical value of $B$ could go even lower for more massive HMNSs with $M_b > 3M_\odot$.
Therefore, the coupling constant $B$ we covered is relatively strong for HMNSs, prolonging the scalarization time and thus explain the strong scalarization behavior.

\subsection{Delayed collapse} \label{sec:collapse}

\begin{figure}
    \centering
    \includegraphics[width=\columnwidth]{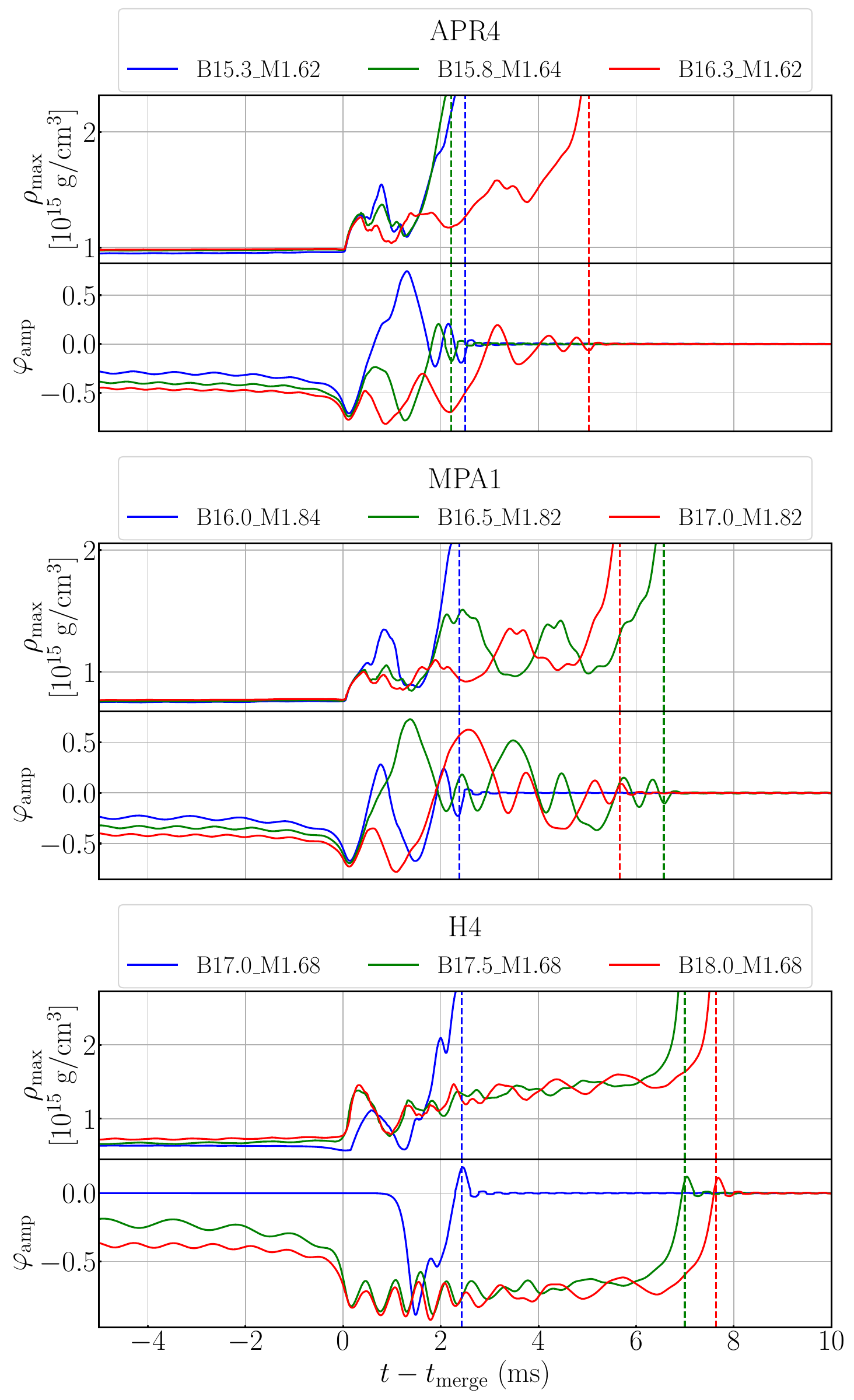}
    \caption{
    Evolution of maximum density $\rho_{\rm max}$ (top) and scalar field amplitude $\varphi_{\rm amp}$ for short-lived HMNS cases.
    The dashed line indicate the collapse time for the corresponding models.
    }
    \label{fig:evo_shortlive}
\end{figure}

When the total mass of merger remnants is slightly below the threshold mass $M_{\rm thr}$, the HMNS survives for a short period of time and then collapses to a black hole after subsequent angular momentum dissipation by the GW emission and angular momentum transport via the gravitational torque associated with non-axisymmetric structure of the HMNS. We classify these delayed collapse models with $\tau_\mathrm{H} < 10~{\rm ms}$ as a short-lived HMNS. We expect that the collapse could be further delayed if the HMNS is spontaneously scalarized since the scalar field will weaken the gravitational force on the surrounding matter.
\cref{fig:evo_shortlive} shows the evolution of $\rho_{\rm max}$ and $\varphi_{\rm amp}$ for short-lived HMNS models. HMNSs with the H4 EOS always remain spontaneously scalarized until the formation of a black hole because of the choice of the relatively strong coupling strength. Then, the descalarization occurs when the black hole is formed and the scalar field is quickly dissipated due to the no-hair theorem.
On the other hand, HMNSs pertaining to the APR4 and MPA1 EOSs undergo descalarization earlier before the black hole formation, leaving an oscillating scalar cloud. These descalarized HMNSs have a mass $>M_{\rm thr}^{\rm GR}$ and yet they still survive for a few ms before forming a black hole. This indicates that the small-amplitude scalar cloud $|\varphi| \lesssim 0.1$ provides a temporal support to stave off the collapse.

\begin{figure}
    \centering
    \includegraphics[width=\columnwidth]{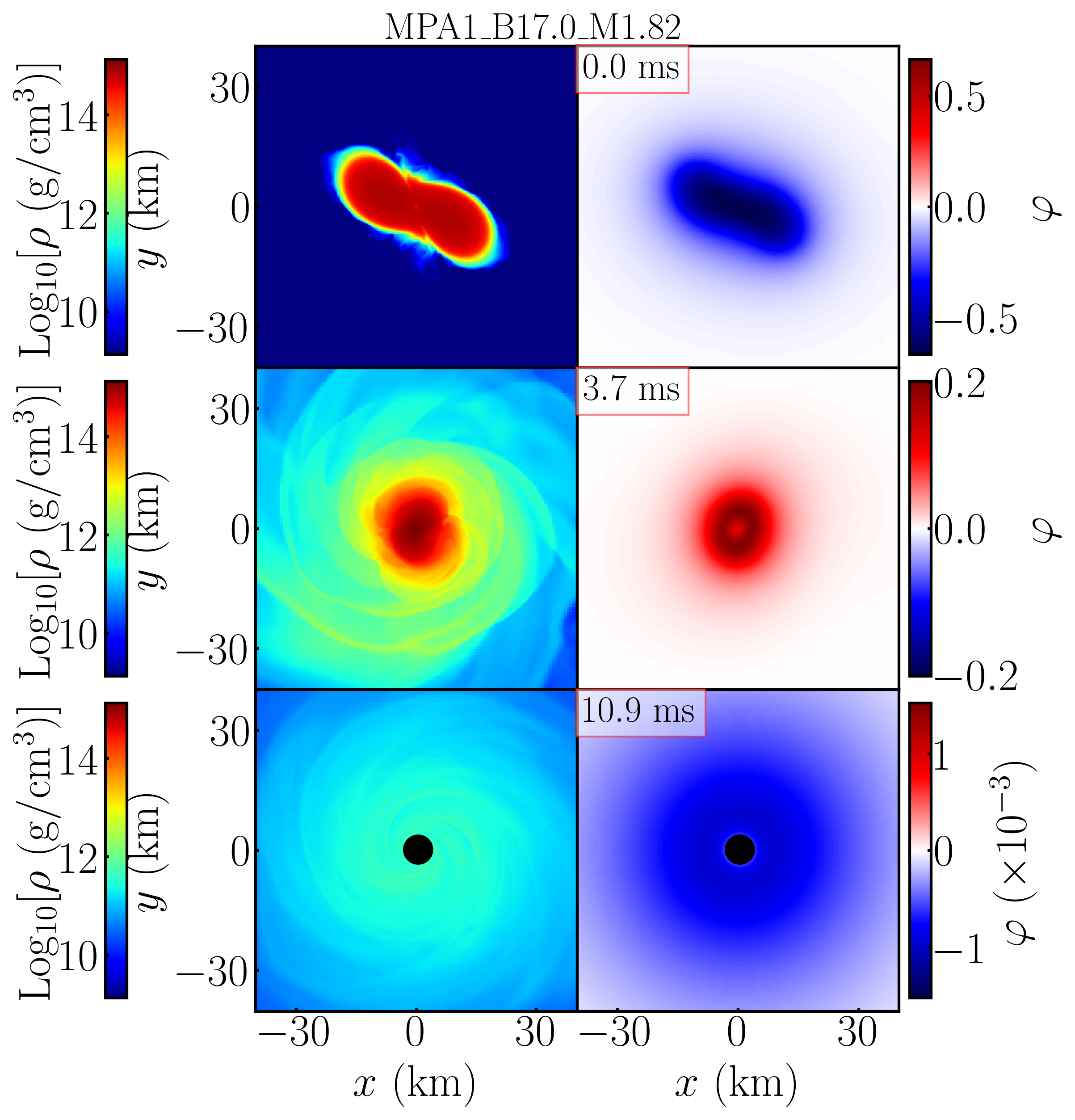}
    \caption{
    Snapshots of rest-mass density $\rho$ and scalar field $\varphi$ on the equatorial plane for a short-lived HMNS model \texttt{MPA1\_B17.0\_M1.82} at the onset of merger (top), before the formation of apparent horizon (middle), and at $10.9~{\rm ms}$ after the onset of merger (bottom). The time after the onset of merger is indicated in the red box and the black filled circles plotted at the bottom panels show the location of the black hole. Notice the varying scale rule for $\varphi$ in different panels.
    }
    \label{fig:shortlive_HMNS}
\end{figure}

Taking one particular model as an example (same one as the red curve for MPA1 in \cref{fig:evo_shortlive}), we find that the evolution of the scalar field and the HMNS in this scenario is visualized in \cref{fig:shortlive_HMNS} through the snapshots of the rest-mass density (left) and scalar field (right) on the equatorial plane. The HMNS descalarizes at $\lesssim2$~ms after the onset of merger and forms a hollow spherical scalar cloud around it (middle), similar to the scalar profile of the descalarized models (cf,~\cref{fig:traceT}).
The scalar cloud delays the collapse of the HMNS until $5.67~{\rm ms}$ after the onset of merger. Eventually, a black hole is formed, which is surrounded by a long-lived quasi-bound state of the scalar cloud with the amplitude of $\sim 10^{-4}$ (bottom) because of the non-zero mass of the scalar field (see more details in \cref{sec:threshold_mass}).

\subsection{Prompt collapse and the threshold mass} \label{sec:threshold_mass}

Shortly after the fully GR BNS merger simulations were feasible, Refs.~\cite{shib03,shib05} showed that there is a mass limit on the BNSs beyond which they immediately collapse into a black hole within a dynamical timescale $\lesssim 1 ~{\rm ms}$. In GR, the threshold mass $M_{\rm thr}^{\rm GR}$ of NSs for which the prompt collapse proceeds has been vastly studied for different EOSs, whereby it was found that this threshold mass varies for different EOSs \cite{shib06,hoto11,baus13,sven19,bern20,kash22}, but is not sensitive to the mass ratio unless the system is appreciably asymmetric as $q<0.7$~\cite{baus21}. The threshold masses for the considered EOSs, APR4, MPA1, and H4, have been found to be 2.825 $M_\odot$, 3.225 $M_\odot$ and 3.125 $M_\odot$, respectively, in \cite{baus20} with GR hydrodynamics simulations under conformal flatness approximation. In addition to dynamical studies, the threshold mass can also be approximately determined by the maximum mass of differentially rotating NSs along a constant angular momentum sequence for a given EOS, i.e., the turning-point criteria is approximately valid to a large extent, provided that the rotational law can be phenomenologically modeled \cite{kapl14,weih18,muha24}.
However in the DEF theory, there could emerge a scalarized branch of equilibrium under the same EOS, angular momentum, and rotational law. The presence of the scalar field in spontaneously scalarized NS will effectively increase the stiffness of the EOS, providing additional support against gravitational collapse and thus the maximum achievable on the scalar branch has been shown to exceed that on the GR-sequence \cite{stay23}. This suggests the existence of HMNSs heavier than the prompt collapse threshold in GR, i.e., the final remnant with mass greater than $M_{\rm thr}$ in GR may not undergo prompt collapse if it is scalarized.

In practical simulations, there is no clear criterion to classify the outcome as the prompt collapse scenario. Some studies \cite{hoto11} used monotonically increasing feature of $\rho_{\rm max}$ after the onset of merger as an indication of the prompt collapse, while some used monotonically decreasing feature of the minimum value of the lapse function, $\alpha_{\rm min}$, toward zero as a criterion \cite{baus21}.
In this study, we employ the minimum lapse function $\alpha_{\rm min}$ as the indicator for the prompt collapse when it decreases monotonically in the merger phase. Although $\alpha_{\rm min}$ is a gauge dependent variable, it directly reflects the geometrical property compared to the maximum rest-mass density $\rho_{\rm max}$ in the DEF framework since the contribution of hydrodynamics is coupled to the scalar field as $\phi^{-1} T_{ab}$ [cf.~\cref{eq:einstein}].
When the remnant undergoes gravitational collapse, the scalar field $|\varphi|$ drops to zero drastically due to the no-hair theorem and causes a small bump in the evolution of the rest-mass density $\rho_{\rm max}$ 

To better resolve the threshold mass for prompt collapse, we increase the grid resolution in binary mass sequence such that the least massive prompt collapse model and the most massive delayed collapse model differ by $\Delta M_b = 0.02 ~M_{\odot}$ in total baryon mass (i.e., $\Delta M_b = 0.01\,M_{\odot}$ for each NS). We define the threshold mass as $M_{\rm thr} := (M_{\rm inf, PC} + M_{\rm inf, SL})/2$ following \cite{baus20} in which $M_{\rm inf, PC}$ and $M_{\rm inf, SL}$ are the ADM masses of least massive prompt collapse model and most massive delayed collapse model at infinite orbital separation, respectively.

\begin{figure}
    \centering
    \includegraphics[width=\columnwidth]{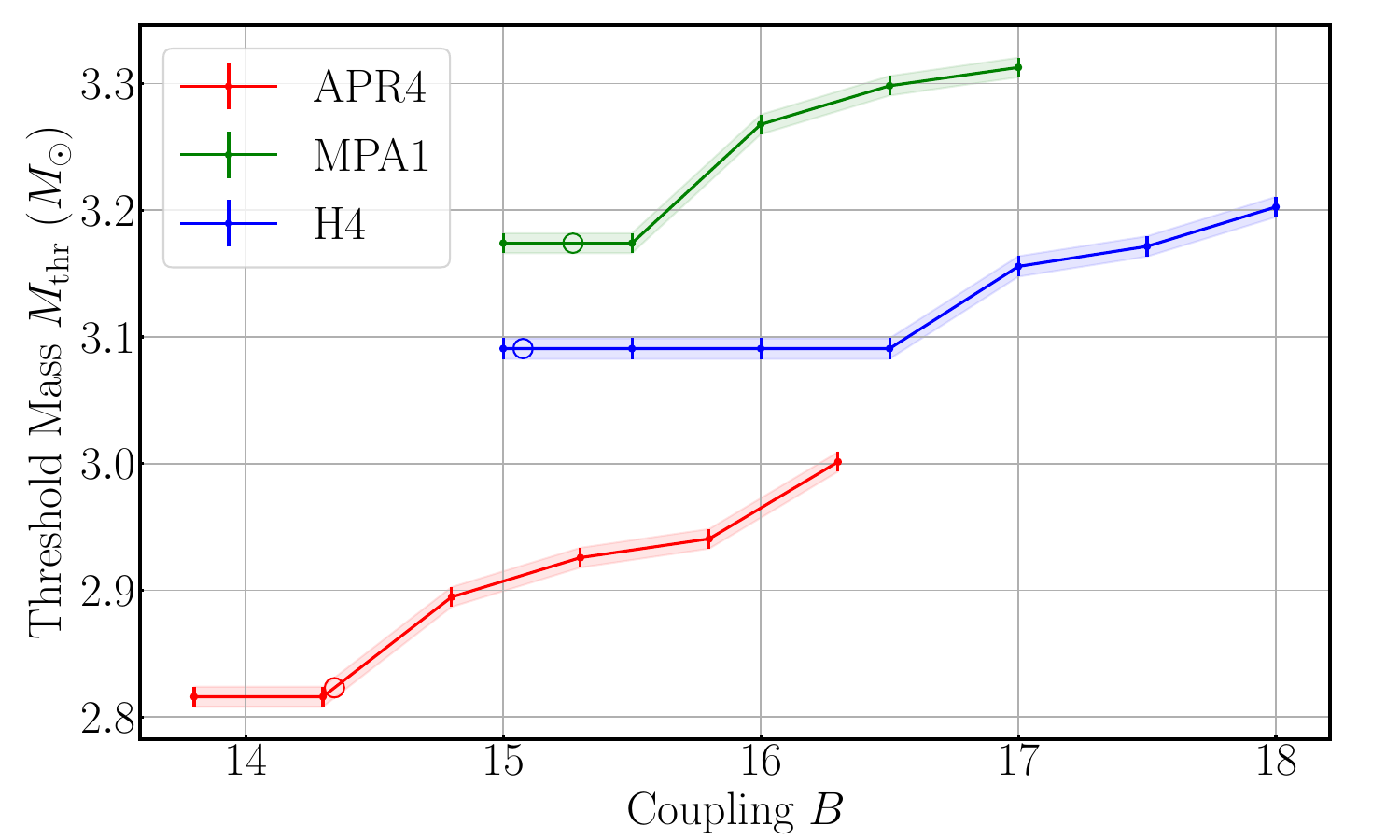}
    \caption{Threshold mass $M_{\rm thr}$ for equal-mass BNSs as a function of the coupling constant $B$ for APR4 (red), MPA1 (green), and H4 (blue) EOSs.
    The circle markers indicate the minimum value of the coupling strength $B^{\rm crit}$ with which spontaneous scalarization is possible for sphereically symmetric NSs.
    The width of each curve reflects the bin size of the mass sampling.
    }
    \label{fig:threshold_mass}
\end{figure}

\cref{fig:threshold_mass} shows the threshold mass of NSs with different values of $B$ for the three EOSs considered. We investigate the dependence of $M_{\rm thr}$ on $B$ until it reaches the minimum coupling strength $B^{\rm crit}$ (circle markers in \cref{fig:threshold_mass}) with which spontaneous scalarization is possible for spherically symmetric NSs as shown in \cref{fig:bdry}. The shaded region indicates the error bar given by $M_{\rm inf, PC}$ and $M_{\rm inf, SL}$. For the weak coupling case $B \lesssim B^{\rm crit}$, the NSs are not scalarized in the inspiral phase, and thus, the contribution of the scalar field is negligible. For this case, the resultant threshold masses are essentially the same as in GR with $M_{\rm thr}^{\rm GR} = 2.816 M_\odot$, $3.174 M_\odot$, and $3.091 M_\odot$ for APR4, MPA1, and H4 EOSs, respectively. Although the obtained threshold masses $M_{\rm thr}^{\rm GR}$ are by $\sim 1\%$ lower than the corresponding values found in \cite{baus20}, this could be due to the systematic error caused by the conformal flatness approximation employed in their study which cannot accurately evolve spacetime with high angular momentum.
This is in agreement with \cite{sven19} in which the obtained $M_{\rm thr}$ is also lower than those in \cite{baus17}.

\begin{figure}
    \centering
    \includegraphics[width=\columnwidth]{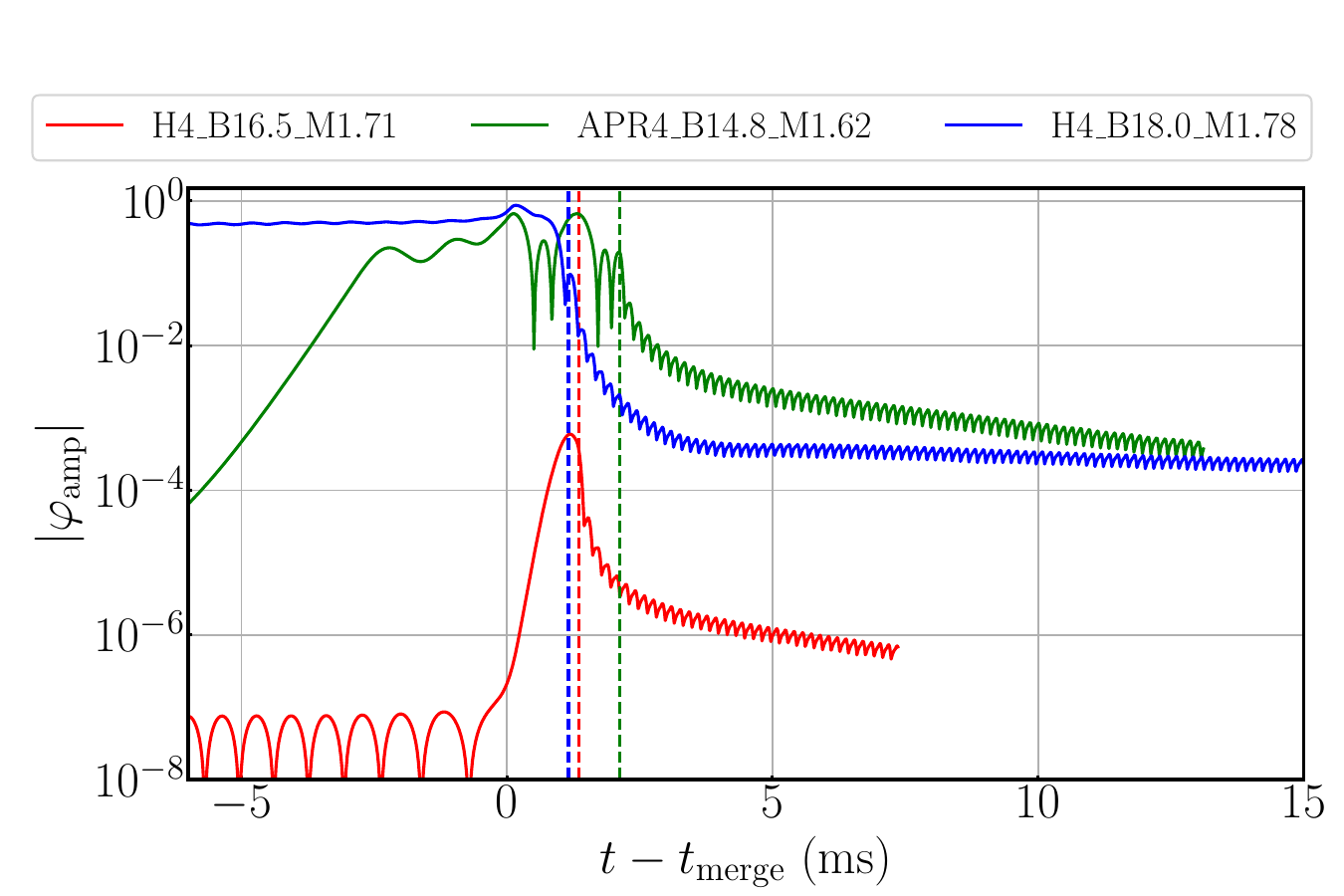}
    \caption{
    Evolution of the maximum scalar-field amplitude $|\varphi_{\rm amp}|$ for three different collapse models.
    The \texttt{H4\_B16.5\_M1.71} (red) and \texttt{H4\_B18.0\_M1.78} (blue) are prompt collapse models while \texttt{APR4\_B14.8\_M1.62} (green) is a delayed collapse model. The colored dotted lines show the collapse times for the corresponding models.
    }
    \label{fig:scalar_pc}
\end{figure}

As the coupling strength $B$ increases, the threshold mass $M_{\rm thr}$ begins to rise when the scalar effect becomes important. Note that whether the threshold mass $M_{\rm thr}$ is modified from GR is determined by scalarization history of the BNS in the inspiral phase. If spontaneous scalarization or dynamical scalarization happens before the merger, the scalar field is large enough to alter the subsequent evolution of the remnant HMNS. Otherwise, even if the final remnant could be potentially scalarized with the associated mass and angular momentum, the scalarization time is longer than the dynamical time of the remnant so that the prompt collapse can happen before the HMNS reaches a state of spontaneous scalarization. This can be found in model \texttt{H4\_B16.5\_M1.71} shown in \cref{fig:scalar_pc} (red) for which the scalar field grows exponentially in the merger phase, hinting a sign of scalarization. However, the remnant undergoes prompt collapse before the scalar field is significantly amplified, and hence, the scalar effect is negligible throughout the evolution process. On the other hand, dynamical scalarization kicks in and gets saturated at 2--3~ms before merger for model \texttt{APR4\_B14.8\_M1.62} (green in \cref{fig:scalar_pc}). Hence, the final remnant is evaded from prompt collapse with total massof $2.887M_\odot$ greater than threshold mass in GR $M_{\rm thr}^{\rm GR}$ of $2.816M_\odot$ because appreciable scalar field is built up in the inspiral phase through the scalarization process.

\begin{figure}
    \centering
    \includegraphics[width=\columnwidth]{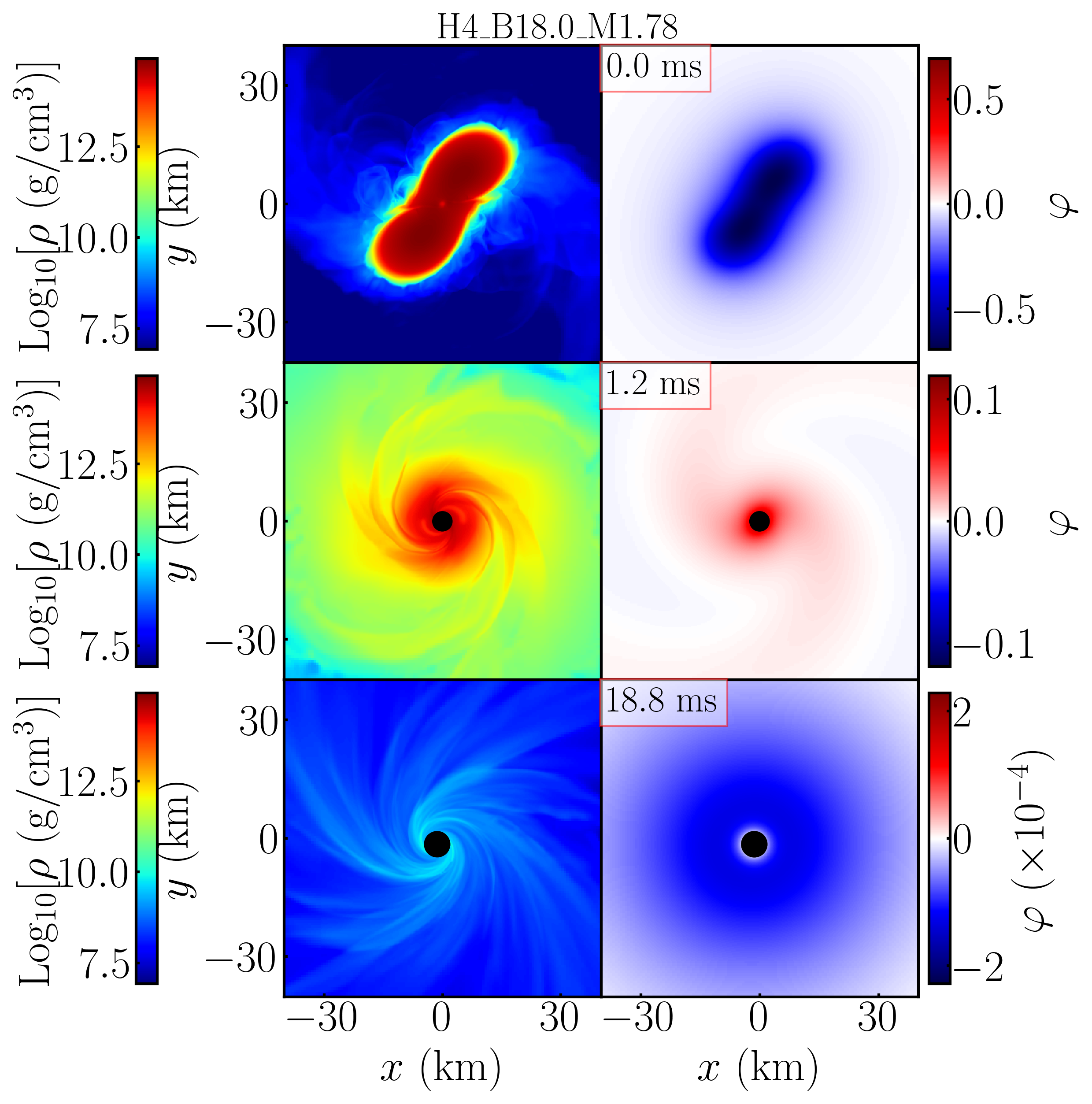}
    \caption{
    Snapshots of rest-mass density $\rho$ and scalar field $\varphi$ on the equatorial plane for a prompt collapse model \texttt{H4\_B18.0\_M1.78} at three different time slices. The time after the onset of merger is indicated in the red box and the black filled circles denote the black hole.
    }
    \label{fig:prompt_collapse}
\end{figure}

As mentioned in \cref{sec:collapse}, after the HMNS collapses, a quasi-bound state of the oscillating scalar cloud will form around the black hole from the fossil scalar field if the system undergoes scalarization beforehand. \cref{fig:scalar_pc} shows that the scalar field for model \texttt{H4\_B18.0\_M1.78} (blue) quickly dissipates most of its energy after the prompt collapse. Nonetheless, a small fraction of the original scalar field remains and settles down to a long-lived oscillating cloud with the amplitude $\sim 10^{-4}$. The final scalar cloud contains dominantly a monopole component as illustrated at the bottom panels of \cref{fig:shortlive_HMNS} and \cref{fig:prompt_collapse}.

\section{Properties of remnants} \label{sec:properties}

\subsection{Dynamical ejecta}

First, we briefly discuss the material ejected from the BNS merger in the DEF theory.

One common method to identify the unbounded fluid element is to use geodesic criteria $u_t \leq -1$ for particles moving on ballistic trajectories~\cite{hoto13b,bern20b,kast15,radi16,nedo22}. We define the total baryon rest-mass $M_{\rm ej}$, total energy $E_{\rm ej}$, and total internal energy $U_{\rm ej}$ of the ejected material by
\begin{align}
    M_{\rm ej}(t) &:= \int_{u_t \leq -1} \rho u^t \sqrt{-g} d^3 x, \\
    E_{\rm ej}(t) &:= \int_{u_t \leq -1} T_{\mu\nu} n^\mu n^\nu \sqrt{\gamma} d^3 x, \\
    U_{\rm ej}(t) &:= \int_{u_t \leq -1} \rho u^t \epsilon \sqrt{-g} d^3 x,
\end{align}
and approximate the kinetic energy $T_{\rm ej}$ as
\begin{align}
    T_{\rm ej}(t) &:= E_{\rm ej} - M_{\rm ej} - U_{\rm ej}.
\end{align}
Assuming that the ejecta has non-relativistic motion, we then estimate the average velocity $v_{\rm ej}$ of it as \cite{hoto13b}
\begin{align}
    v_{\rm ej}(t) &:= \sqrt{\frac{2 T_{\rm ej}}{M_{\rm ej}}}.
\end{align}
However, the influence of gravitational potential still remains in $T_{\rm ej}$ as evaluated within the computation domain $\lesssim 7500~{\rm km}$, hence overestimating the ejecta velocity. We therefore further estimate the extrapolated velocity $v_{\rm ej, ex}$ following \cite{haya21,chen24} as
\begin{align}
    v_{\rm ej, ex}(t) := \sqrt{v_{\rm ej}^2 - 2 \frac{M_{\rm inf}}{{v_{\rm ej} \times (t-t_{\rm merge})}}},
\end{align}
where $v_{\rm ej}$ is evaluated at time $t$. In this paper we define the mass $M_{\rm dyn}$ and the average velocity $v_{\rm dyn}$ of unbounded dynamical ejecta at 10~ms after the onset of merger from $M_{\rm ej}$ and $v_{\rm ej, ex}$, respectively.
Note that due to the residual eccentricity $e\sim 10^{-2}$ in our simulations and limited grid resolution, the total mass of the ejected material could be altered by $\mathcal{O}(10\%)$ compared to circular orbits \cite{fouc24}.

\begin{figure}
    \centering
    \includegraphics[width=\columnwidth]{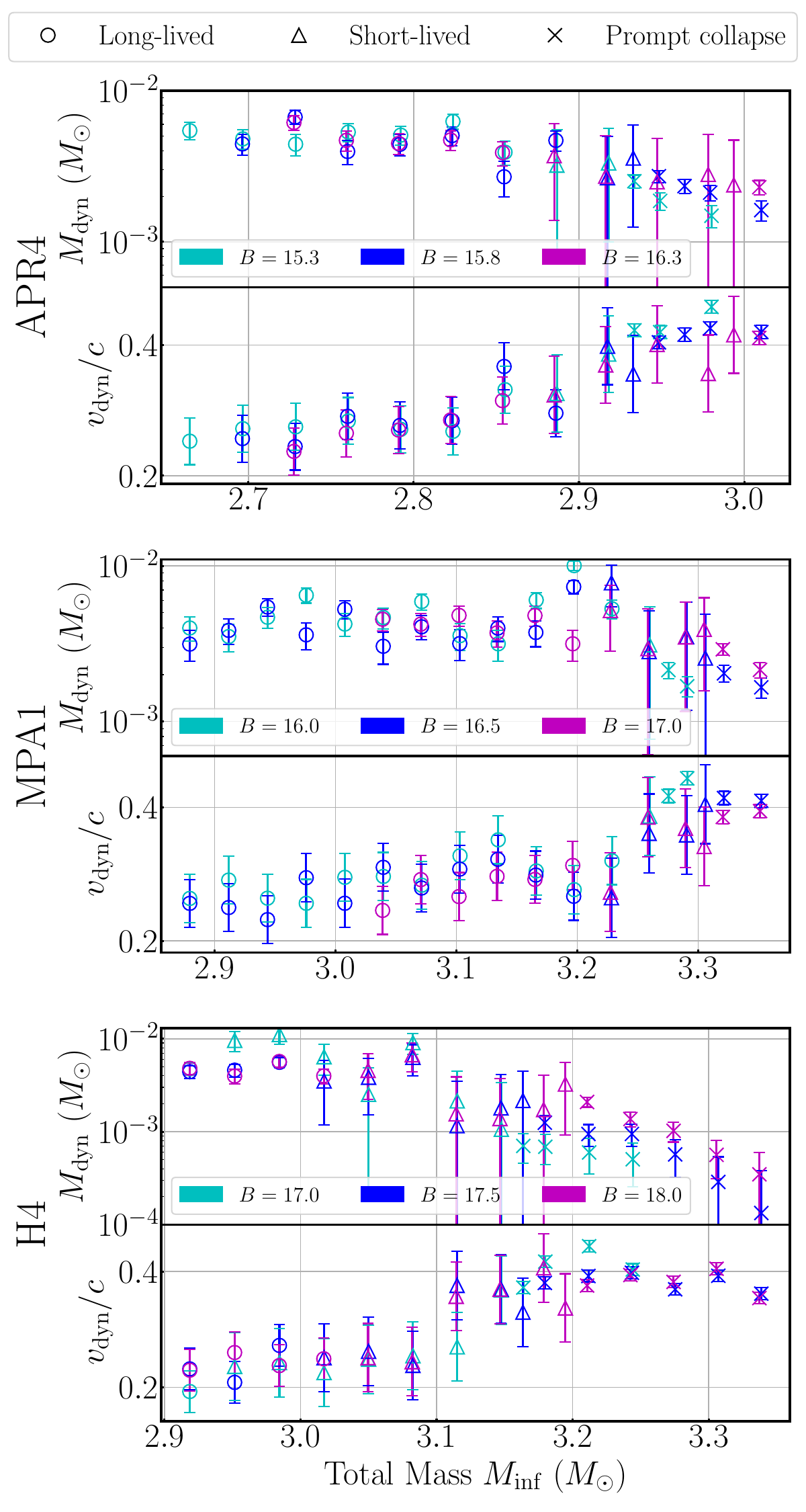}
    \caption{
    Dynamical ejecta mass ($M_{\rm dyn}$) and extrapolated average velocity ($v_{\rm dyn}$) as functions of mass $M_{\rm inf}$ for all the simulated BNS models. Each panel refers to a given EOS, while different coupling strengths $B$ are distinguished by different colors. The circle, triangle, and cross markers represent long-lived HMNS, short-lived HMNS, and prompt collapse models, respectively.
    The error bars are estimated from the convergence test shown in \cref{conv.test}.
    }
    \label{fig:ejecta_prop}
\end{figure}

\cref{fig:ejecta_prop} summaries the total mass $M_{\rm dyn}$ and extrapolated average velocity $v_{\rm dyn}$ of the dynamical ejecta. The circle, triangle, and cross markers represent long-lived HMNS, short-lived HMNS and prompt collapse models, respectively.
The error bars are estimated by the convergence test for long-lived HMNSs, short-lived HMNSs and prompt collapse cases: see \cref{conv.test}.
Since the collapse time is very sensitive to the grid resolution in the short-lived HMNS formation and hence alters the final ejecta properties, the corresponding error bar is much larger than the other two cases.
The ejecta mass $M_{\rm dyn}$ falls in the range of $10^{-3}$--$10^{-2}~M_\odot$ depending on the EOS for the long-lived HMNS formation case with the average velocity $v_{\rm dyn} \sim 0.2c$--$0.3c$. The ejecta mass is found to be often very low as $\alt 10^{-3}~M_\odot$ for the prompt collapse case (in particular for the H4 EOS) due to inefficient time for outward angular momentum transport. For the APR4 and MPA1 EOSs, the ejecta mass is not extremently low as $\agt 10^{-3}M_\odot$. The reason for this is that we pay particular attention to the BNS mass which is close to the threshold of the prompt collapse, and thus, shock heating effects at the merger induce a certain amount of the dynamical mass ejection. For these models the ejecta velocity becomes fairly high $0.3$--$0.4c$ because the shock heating is the dominant source of the dynamical mass ejection.

We find that the ejecta properties are determined primarily by the lifetime of HMNSs while the scalar effect is minor for the long-lived/short-lived HMNS formation case. This is reasonable because the dynamical ejecta quickly escapes the Compton wavelength $\comp \approx 15~{\rm km}$ of the scalar field, and hence, the ejecta evolution is not significantly influenced by the scalar effect. This picture may change for lower values of $m_\phi$, while observationally allowed values of $B$ will be further bounded to lower values. 

\subsection{Black hole and disk}
For models that undergoes gravitational collapse to a black hole, we estimate the parameters of the black hole from the equatorial circumferential radius $C_{\rm e}$ and the area $A_{\rm AH}$ of the apparent horizon by assuming that the spacetime is approximately stationary with negligible effect from the matter. The black hole's mass $M_{\rm BH}$ and dimensionless spin parameter $\chi_{\rm BH}$ can be approximately computed via \cite{shib09}
\begin{align}
    M_{\rm BH} &= \frac{C_{\rm e}}{4 \pi}, \\
    \chi_{\rm BH} &= \sqrt{ 1 - \left( \frac{A_{\rm AH}}{8\pi M_{\rm BH}^2} - 1 \right)^2},
\end{align}
respectively.
Here, we evaluate $M_{\rm BH}$ and $\chi_{\rm BH}$ at 10~ms after the apparent horizon is formed.
The total bounded baryon rest mass outside the apparent horizon is determined via
\begin{align}
    M_{\rm disk}(t) := \int_{r > r_{\rm AH}} \rho u^t \sqrt{-g} d^3 x - M_{\rm ej}(t),
\end{align}
with $r_{\rm AH} = r_{\rm AH}(\theta, \phi)$ being the coordinate radius of the apparent horizon. We also refer to the final disk mass $M_{\rm disk, 0}$ as $M_{\rm disk}(t-t_{\rm AH}=10\,{\rm ms})$, where we recall that $t_{\rm AH}$ is the first formation time of the apparent horizon.

\begin{figure}
    \centering
    \includegraphics[width=\columnwidth]{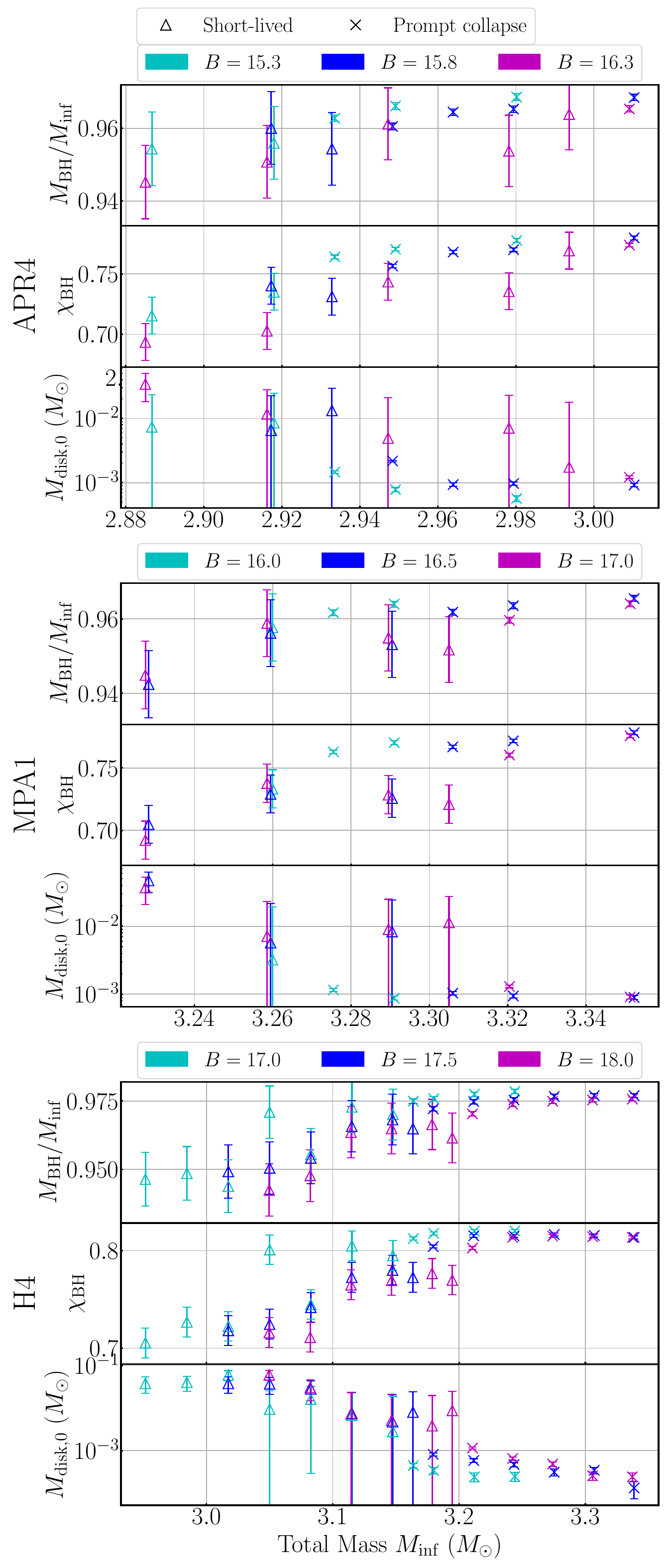}
    \caption{Summary of final black hole and disk properties for short-lived HMNS formation (dot markers) and prompt collapse (cross markers) cases with APR4 (top), MPA1 (middle), and H4 (bottom) EOSs.
    For each EOS subplot, the black hole mass scaled by the total mass $M_{\rm BH}/M_{\rm inf}$ (top), dimensionless spin parameter $\chi_{\rm BH}$ (middle), 
    and the final disk mass $M_{\rm disk,0}$ (bottom) are shown.
    The error bars are estimated from the convergence test shown in \cref{conv.test}.
    }
    \label{fig:bh_torus}
\end{figure}

We summarize the properties of the black hole and disk in \cref{fig:bh_torus} for short-lived HMNS formation and prompt collapse models. For the prompt collapse models (cross markers), the remnant disk mass is significantly suppressed with $M_{\rm disk,0}\lesssim 10^{-3}~M_\odot$ due to the insufficient time for angular momentum to be transported outwards and hence most of the matter falls into the BH as shown by the relatively high $M_{\rm BH}/M_{\rm inf}$ factor and dimensionless spin parameter $\chi_{\rm BH}$. Nonetheless, the dynamical timescale for the remnant to collapse to a black hole is slightly extended for larger values of $B$ due to the decrease in compactness of isolated NSs. For example, the lifetime $\Tns$ rises from 0.83~ms in \texttt{H4\_B17.0\_M1.80} to 1.01~ms in \texttt{H4\_B18.0\_M1.80} as the coupling strength $B$ is increased from 17 to 18. As a result, more matter remains outside the black hole, yielding a slight decrease in $M_{\rm BH}$ and $\chi_{\rm BH}$.

\begin{figure}
    \centering
    \includegraphics[width=\columnwidth]{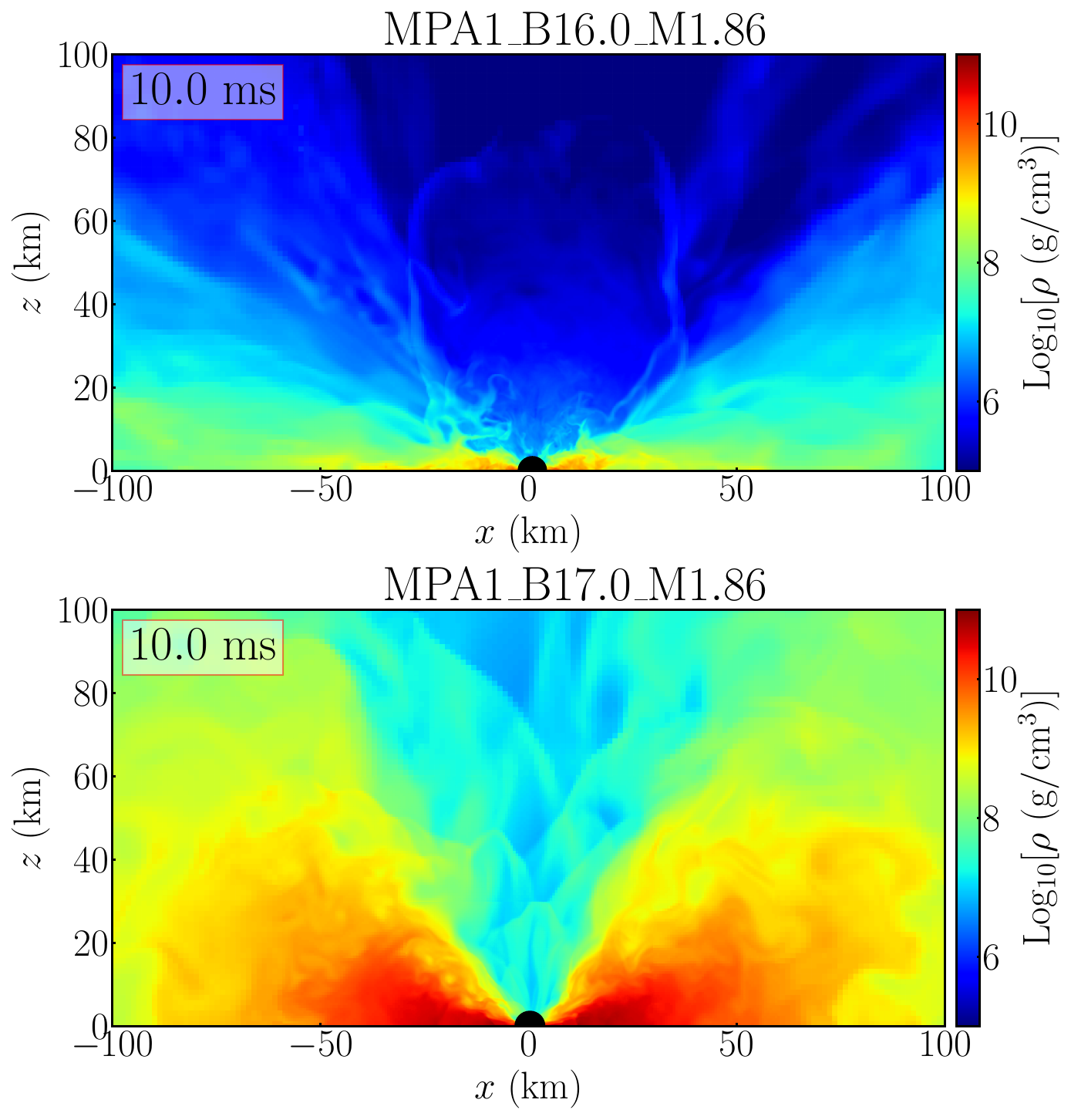}
    \caption{
    Snapshots of rest-mass density $\rho$ on the $x$-$z$ plane for a prompt collapse model
    \texttt{MPA1\_B16.0\_M1.86} (top) and a short-lived HMNS formation model \texttt{MPA1\_B17.0\_M1.86} (bottom) at 10~ms after the formation of apparent horizon. The black filled circles at the center denote the black hole.
    }
    \label{fig:snapshot_disk}
\end{figure}

For the short-lived HMNS formation case, the disk mass is much higher than for the prompt collapse case, and typically falls in the range of $\sim 10^{-2}-10^{-3}~M_\odot$. Simultaneously, the resultant black hole mass and spin are  lower. This result is consistent with that found in GR hydrodynamics; the lifetime of the HMNS primarily determines the final disk mass in the case of equal-mass BNSs. Since $M_{\rm thr}$ could be modified for large enough values of $B$ in the DEF theory, the disk mass could be significantly modified compared to in GR with the same value of  $M_{\rm inf}$. 

\cref{fig:snapshot_disk} shows the snapshots of the disk on the $x$-$z$ plane at 10~ms after the formation of the apparent horizon for \texttt{MPA1\_B16.0\_M1.86} and \texttt{MPA1\_B17.0\_M1.86}. Despite of their similar masses $M_{\rm inf}$ ($\Delta M_{\rm inf} < 0.002 M_\odot$), the short-lived HMNS formation model \texttt{MPA1\_B17.0\_M1.86} has a thick torus with mass $M_{\rm disk,0} = 7.3\times 10^{-3} M_\odot$ outside the horizon, while only a thin disk with tiny mass $M_{\rm disk,0} = 5.1\times 10^{-4}M_\odot$ remains in the black hole's proximity for the prompt collapse model \texttt{MPA1\_B16.0\_M1.86}.

\begin{figure}
    \centering
    \includegraphics[width=\columnwidth]{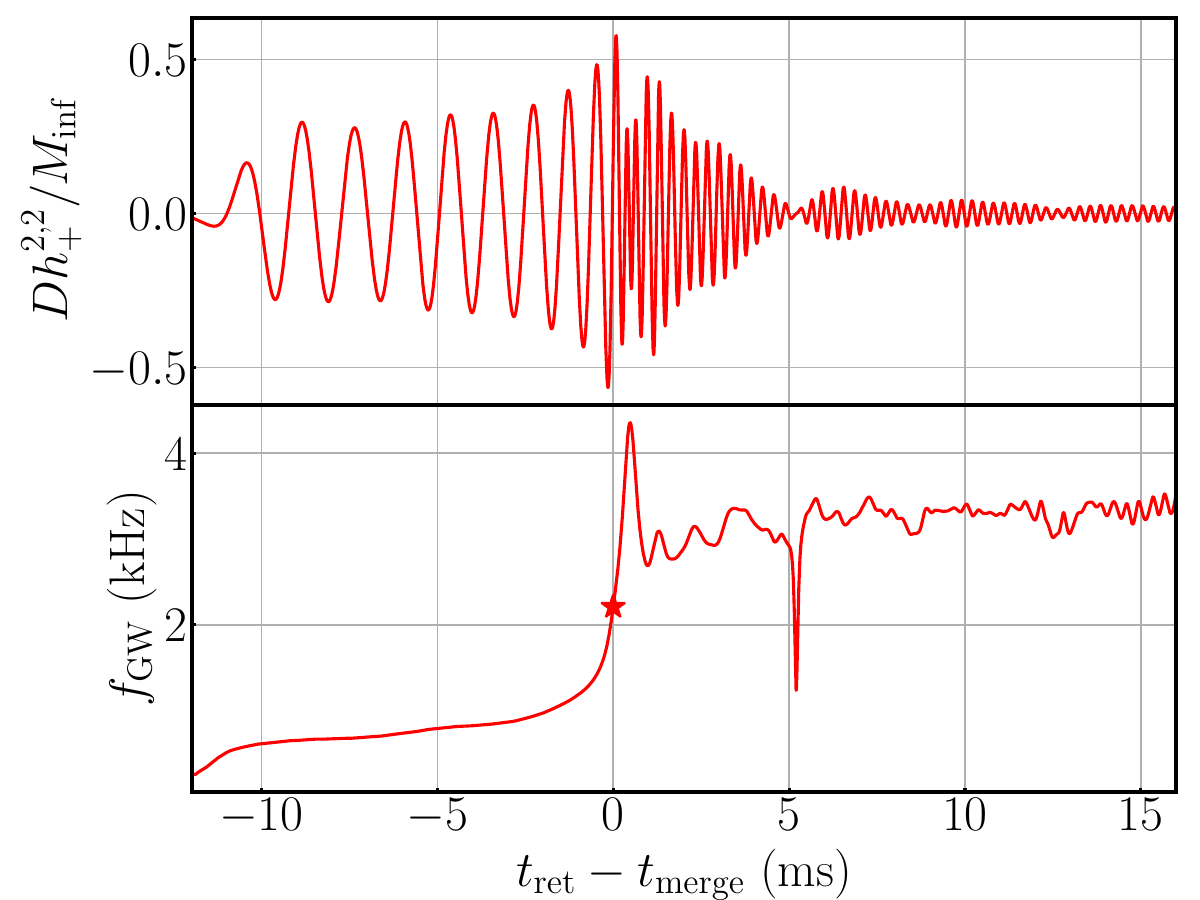}
    \caption{
    GWs emitted by \texttt{APR4\_B15.8\_M1.56}. The top panel shows the plus \addal{$h_+$ (red)}
    polarization of GWs normalized by the extraction radius $D=480~M_\odot$ and initial ADM mass of BNSs, \addal{$M_{\rm inf}$}, as a function of retarded time \addal{$t_{\rm ret}-t_{\rm merge}$}. The bottom panel shows the instantaneous GW frequency $f_{\rm GW}$. The red star marker indicates the merge frequency $f_{\rm merge}$.
    }
    \label{fig:GW_APR4}
\end{figure}

\subsection{Characteristics of gravitational waves from descalarized HMNS}

\begin{figure}
    \centering
    \includegraphics[width=\columnwidth]{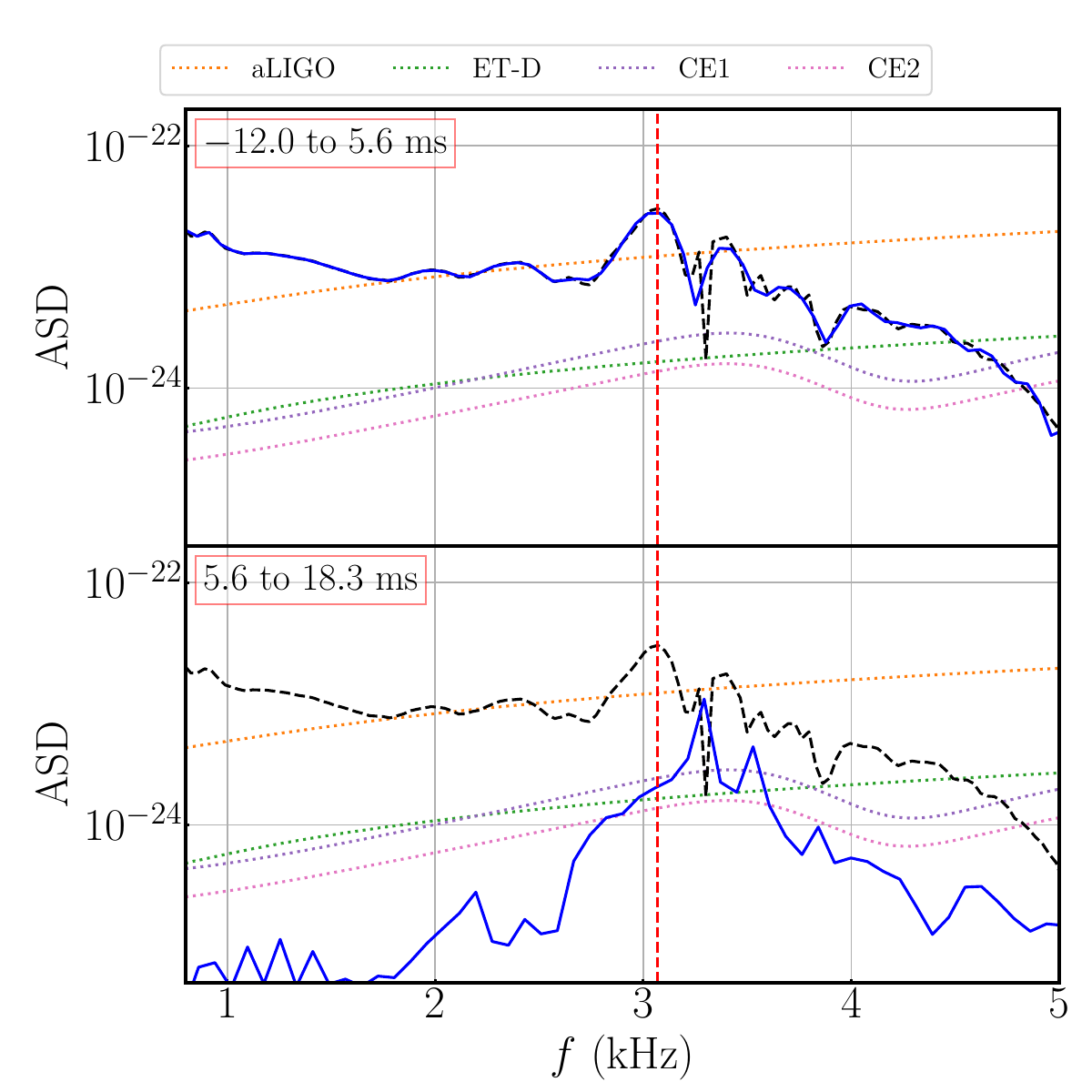}
    \caption{
    ASD $\tilde h \sqrt{f}$ (${\rm Hz}^{-1/2}$) of \texttt{APR4\_B15.8\_M1.56} at a distance of 50~Mpc. The black dashed curve indicates the ASD of the the whole waveform and the vertical red dashed line indicates the $f_{\rm 2, peak}$. 
    The blue line in the upper and lower panels show respectively the ASD of the waveform before and after the onset of descalarization ($5.6~{\rm ms}$ after merge).
    }
    \label{fig:GW_post_APR4}
\end{figure}

In this paper, we focus on the discussion for a property of post-merger waveforms that is special to the scenarios involving a descalarization, while leaving more extensive investigation about other scenarios to future paper (Lam et al., in preparation). Taking model \texttt{APR4\_B15.8\_M1.56} as an example, \cref{fig:GW_APR4} shows the plus polarization (top) and simultaneous frequency [\cref{eq:instan_fgw}; bottom] of the GW signal. We denote the instantaneous frequency at the onset of merger at which the absolute amplitude $|h|$ reaches its maximum as $f_{\rm merge}$, which is sometimes denoted as $f_{\rm peak}$ or $f_{\rm 2, max}$ in the literature. 
We also define $f_{\rm 2, peak}$ as the frequency as the dominant peak in the Fourier spectrum of $h_{\rm eff}$ in the post-merger phase, which is attributed to the $l=m=2$ mode of the HMNS \cite{shib05b,ster11,hoto11,taka14,taka15,fouc16}. The acceleration spectral density (ASD) $\tilde h \sqrt{f}~({\rm Hz}^{-1/2})$ is plotted in \cref{fig:GW_post_APR4} for this model assuming a source distance of 50~Mpc. Since the HMNS in this model undergoes descalarization at 5.6~ms after the onset of merger,
we perform the Fourier analysis of the waveform within two different time segments before and after descalarization indicated by the solid blue curves on the top and bottom panels in \cref{fig:GW_post_APR4}, respectively, while the spectrum of the whole waveform is shown by the black dashed curve.
By comparing the spectrum of the whole waveform to that of the two time windows, we find that the $f_{\rm 2, peak}$ is determined primarily by the state of the HMNS at a few ms after the onset of merger. In the later time window, we find an up-wind shift in $f_{\rm 2, peak}$ after the descalarization since the compactness of the HMNS increase during this process. Both the increased compactness and the higher $f_{\rm 2, peak}$ are similar characteristics of the GW signature shared with the influence of a phase transition from confined hadronic matter to deconfined quark matter (e.g., \cite{weih20,blac20,baus19a,baus19b}).

\begin{figure}
    \centering
    \includegraphics[width=\columnwidth]{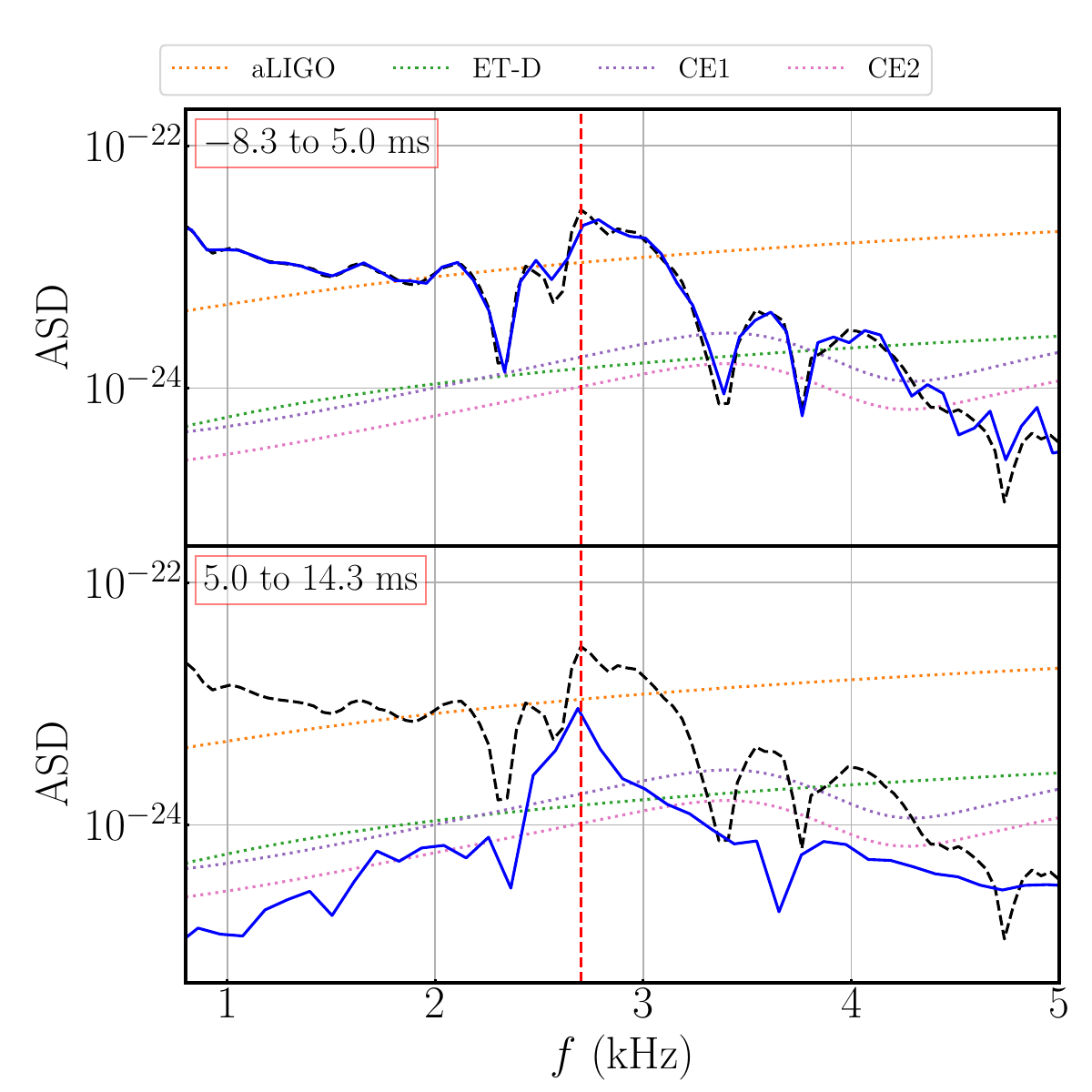}
    \caption{
    ASD $2 \tilde h \sqrt{f}$ (${\rm Hz}^{-1/2}$) of \texttt{H4\_B18\_M1.64} at a distance of 50\,Mpc. The black dashed curvee indicates the ASD of the the whole waveform and the vertical red dashed line indicates the $f_{\rm 2, peak}$.
    The blue line in the upper and lower panels show, respectively, the ASD of the waveform before and after $t-t_{\rm merge} = 5~{\rm ms.}$
    }
    \label{fig:GW_post_H4}
\end{figure}

For comparison, we show in \cref{fig:GW_post_H4} the ASD in two time segments separated by 5~ms after the onset of merger for model \texttt{H4\_B18.0\_M1.64}, whereas the remnant HMNS remains scalarization in the post-merger phase. The $f_{\rm 2, peak}$ does not shift in the absence of a state transition in the HMNS throughout the post-merger phase, which verifies that the shift in $f_{\rm 2, peak}$ is indeed caused by the state transition of descalarization.

%%%%%%%%%%%%%%%%%%%%%%%%%%%%%%%%%%
\section{Summary and Discussion} \label{sec5}
%%%%%%%%%%%%%%%%%%%%%%%%%%%%%%%%%%

We performed numerical relativity simulations to study the properties of post-merger remnants and GW emission from BNS mergers in the DEF theory with a massive scalar field. We focused on a canonical scalar mass of $m_\phi = 1.33 \times 10^{-11}~{\rm eV}$ suggested in \cite{kuan23} to explore a wide range of NS mass and coupling strength $B$ for the APR4, MPA1, and H4 EOSs. In the framework of the DEF theory, a scalar cloud can be induced in NSs and HMNSs by spontaneous scalarization or through dynamical scalarization in the binary system. This additional scalar field modifies the classic picture of BNS post-merger remnants. In the presence of scalarization, the lifetime of the HMNSs is prolonged due to the extra support from the scalar field.
This raises the threshold mass for the prompt collapse by $0.1$--$0.2~M_\odot$, which depends on the EOS (\cref{fig:threshold_mass}). 

For lower BNSs from which a long-lived HMNS is formed, the excited scalar field also changes its dynamics from GR one. We find that the remnant can undergo descalarization if the maximum density reaches a certain critical value to become ultrarelativistic (\cref{fig:traceT}), either due to the merger or subsequent post-merger evolution by the GW emission and the angular momentum redistribution via gravitational torque associated with the non-axisymmetric structure of the remnant. Afterward, an oscillating scalar cloud remains in the vicinity of the descalarized HMNS, and lasts over 10~ms after descalarization with appreciable amplitude $\Delta \varphi \lesssim 0.1$ (\cref{fig:longlive_evo,fig:marginal_evo}) instead of rapidly dissipating away as that would happen for a massless scalar field.
Not only in a descalarized HMNS can we observe a long-lived $\phi$-mode. Even for HMNSs that remain scalarized to the end of the simulation, the $\phi-$mode 
excited during merger is exhibited (\cref{fig:longlive_evo}), and helps enhancing a quasi-radial oscillation in the HMNS. Such a long-lived scalar cloud can also be found even after the HMNS collapses to a black hole while with much smaller amplitude (\cref{fig:scalar_pc}).

The scalar field alters the lifetime of HMNSs (\cref{fig:lifetime}), which in turn modifies the dynamical ejecta mass and disk mass. This may give a different kilonova signature from the GR prediction for a system with the same mass. We also observe an upward shift in $f_{\rm 2, peak}$ frequency in post-merger GW signal due to the transition in the HMNS's state caused by descalarization (\cref{fig:GW_post_APR4}),
which assembles the characteristics of the EOS phase transition when deconfined quark matter reveals. The result for more detailed analysis of gravitational waveforms and their spectra will be presented in a separate paper.

%%%%%%%%%%%%%%%%%%%%%%%%%%%%%%%%%%
\section*{Acknowledgement}
%%%%%%%%%%%%%%%%%%%%%%%%%%%%%%%%%%
We thank the member of the Computational Relativistic Astophysics department of the Max Planck Institute for Gravitational Physics for ueful discussion. Numerical computation was performed on the clusters Sakura and Cobra at the Max Planck Computing and Data Facility. This work was in part supported by Grant-in-Aid for Scientific Research (grant Nos.~20H00158, 23H04900, and 23K25869) of Japanese MEXT/JSPS.

\appendix 

\section{Convergence test}\label{conv.test}

We summarize the details of numerical setup used in the simulations in \cref{tab:numerical_setup}.
We adopt $N=94$ as the standard resolution throughout this paper.

\begin{table}
    \centering
    \caption{Numerical setups for the simulations.
    The grid number for covering one positive direction $(N)$,
    the grid spacing in the finest refinement level $(\Delta x)$, the total size of computation domain $[-L, L]$, total number of moving boxes $(n_{\rm fix})$ and fixed (non-moving) boxes $(n_{\rm fix})$, total number of refinement depths $(d)$ and the extraction radius $(r_{\rm ex})$.}
    \begin{tabular}{
        >{\centering}p{0.10\columnwidth}
	>{\centering}p{0.14\columnwidth}
        >{\centering}p{0.20\columnwidth}
	>{\centering}p{0.08\columnwidth}
        >{\centering}p{0.08\columnwidth}
        >{\centering}p{0.08\columnwidth}
        >{\centering\arraybackslash}p{0.18\columnwidth}
    }
    \hline\hline \\[-.8em]
    N & $\Delta x$ $(m)$ & L $(10^{6} m)$ & $n_{\rm mv}$ & $n_{\rm fix}$ & $d$ & $r_{\rm ex}$ $(km)$ \\
    \hline
    78 & 189 & 7.56 & 8 & 6 & 10 & 709 \\
    \hline
    94 & 157 & 7.56 & 8 & 6 & 10 & 709 \\
    \hline
    110 & 134 & 7.56 & 8 & 6 & 10 & 709 \\
    \hline\hline
    \end{tabular}
    \label{tab:numerical_setup} 
\end{table}

\begin{figure}
    \centering
    \subfloat[Long-lived HMNS \texttt{MPA1\_B16.5\_M1.76}.
        \label{subfig:conv_176}]{%
        \includegraphics[width=\columnwidth]{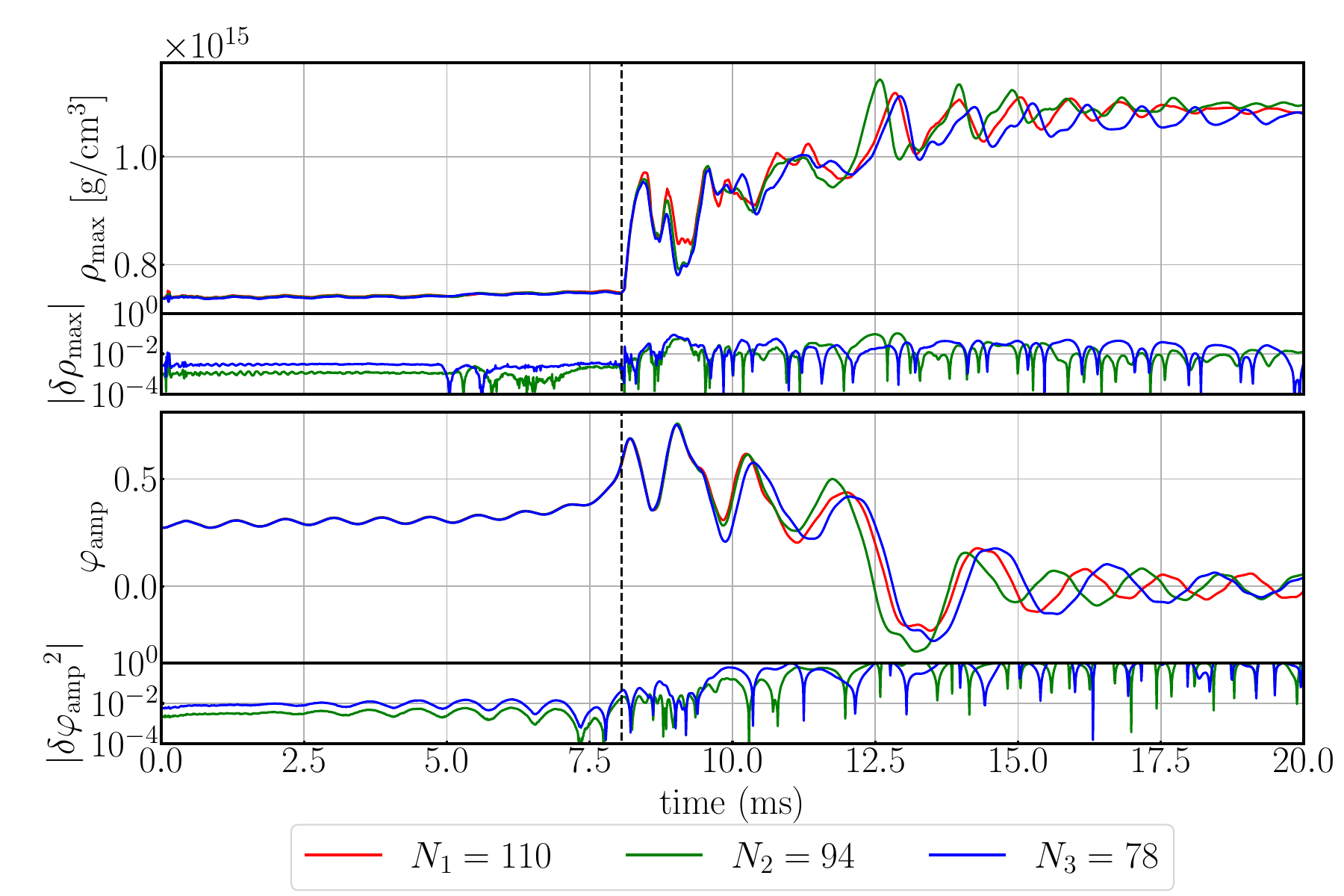}%
    }\hfill
    \subfloat[Short-lived HMNS\texttt{MPA1\_B16.5\_M1.82}.
        \label{subfig:conv_182}]{%
        \includegraphics[width=\columnwidth]{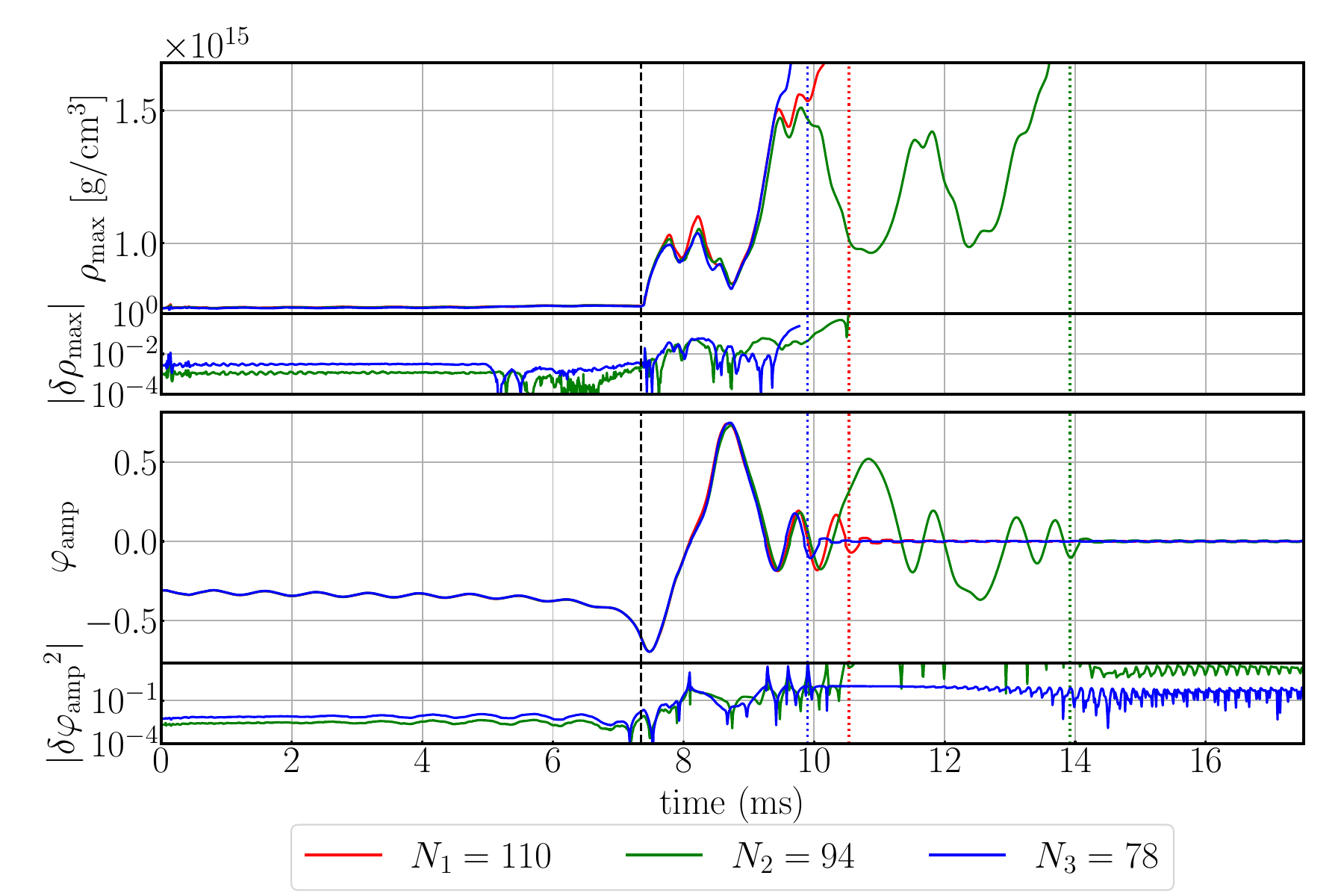}%
    }\hfill
    \subfloat[Prompt collapse \texttt{MPA1\_B16.5\_M1.88}.
        \label{subfig:conv_188}]{%
        \includegraphics[width=\columnwidth]{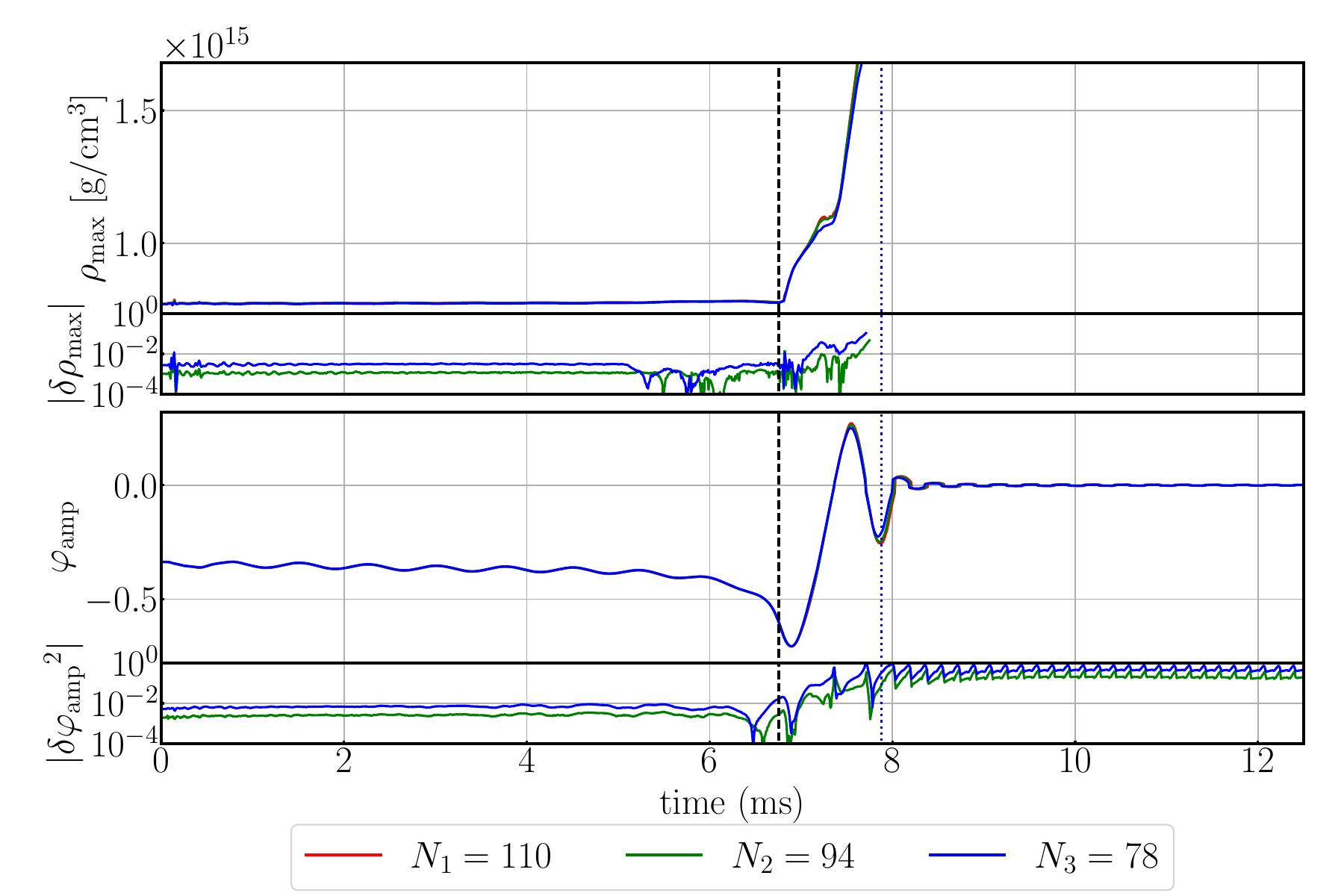}%
    }
    \caption{Convergence test for three different models.
        In each subplot, the upper panel shows the evolution of maximum density $\rho_{\rm max}$ with resolution $N_1 = 110$ (red),
        $N_2 = 94$ (green) and
        $N_3 = 78$ (blue),
        together with the relative error $|\delta \rho_{\rm max}|:=|\rho_{\rm max}/\rho_{1, {\rm max}} - 1|$ 
        in $N_{2,3}$ with respect to the highest resolution $\rho_{1, {\rm max}}$.
        The lower panel shows the evolution of scalar field $\varphi_{\rm amp}$ and
        the relative error 
        $|\delta \varphi_{\rm amp}{}^2|:=
        |\varphi_{\rm amp}{}^2/\varphi_{1, {\rm amp}}{}^2 -1|$ in the corresponding resolutions.
        The black dashed line shows the merger time in $N_1$ while the colored dotted lines in (b) and (c) show the collapse time in different resolutions respectively.
    }
    \label{fig:conv}
\end{figure}

\cref{fig:conv} shows a result of the convergence test considering models of
long-lived HMNSs, \texttt{MPA1\_B16.5\_M1.76} (\cref{subfig:conv_176}), short-lived HMNSs, \texttt{MPA1\_B16.5\_M1.82} (\cref{subfig:conv_182}), and prompt collapse, \texttt{MPA1\_B16.5\_M1.88} (\cref{subfig:conv_188}) with three different grid-resolutions as $N = (110, 94, 78)$.
We obtain convergent result in the inspiral phase
, while the poor resolution in the post-merger phase become notable in the presence of shocks.
In particular for the short-lived HMNS formation model, \texttt{MPA1\_B16.5\_M1.82} (\cref{subfig:conv_182}), $\rho_{\rm max}$ and $\varphi_{\rm amp}$ deviate significantly at 2\,ms after the onset of merger with non-converging collapse time since the evolution of the marginally stable HMNS is extremely sensitive to the grid resolution. Nonetheless, we find consistent evolution of $\rho_{\rm max}$ and $\varphi_{\rm amp}$ for the cases of long-lived HMNS formation model, \texttt{MPA1\_B16.5\_M1.76} (\cref{subfig:conv_176}), and prompt collapse model,  \texttt{MPA1\_B16.5\_M1.88} (\cref{subfig:conv_188}).
In addition, the descalarization time $\Tsc$ and the collapse time in \texttt{MPA1\_B16.5\_M1.76} and \texttt{MPA1\_B16.5\_M1.88}, respectively, have a good convergence. This indicates that the standard resolution $\Delta x = 157 ~{\rm m}$ we employed throughout this paper is acceptable to explore the scenarios of long-lived HMNS formation and prompt collapse.

\begin{table}
    \centering
    \caption{Errors of remnant properties for long-lived HMNSs, short-lived HMNSs and prompt collapse cases.}
    \begin{tabular}{
        >{\centering}p{0.20\columnwidth}
	>{\centering}p{0.14\columnwidth}
        >{\centering}p{0.14\columnwidth}
	>{\centering}p{0.14\columnwidth}
        >{\centering}p{0.14\columnwidth}
        >{\centering\arraybackslash}p{0.14\columnwidth}
    }
    \hline\hline \\[-.8em]
    Models & $\Delta M_{\rm dyn}$ $10^{-4} M_\odot$& $\Delta v_{\rm dyn}$ $10^{-2}$ & $\Delta M_{\rm disk}$ $M_\odot$ & $\Delta M_{\rm BH}$ $10^{-3} M_\odot$ & $\Delta \chi_{\rm BH}$ $10^{-3}$ \\
    \hline
    Long-lived HMNS & 2.5 & 1.0 & - & - & - \\
    \hline
    Short-lived HMNS & 23.3 & 5.9 & 5.9 $\times 10^{-5}$ & 2.7 & 1.5 \\
    \hline
    Prompt Collapse & 7.1 & 3.6 & 1.6 $\times 10^{-2}$ & 29.2 & 15.2 \\
    \hline\hline 
    \end{tabular}
    \label{tab:error} 
\end{table}
We estimate the errors of dynamical ejecta mass $M_{\rm dyn}$ and velocity $v_{\rm dyn}$, remnant disk mass $M_{\rm disk}$ and black hole parameters $M_{\rm BH}$, $\chi_{\rm BH}$ by their difference under the three resolutions considered,
which are given by \cref{tab:error}.

\section{List of the selected Models}\label{modellist}

In Tables~\ref{tab:outcomes_APR4}, \ref{tab:outcomes_MPA1}, \ref{tab:outcomes_H4}, we summarize the outcomes for all the models considered in this paper. 
%%%%%%%%%%%%%%%%%%%%%%%%%%%%%%%%%%%%%%%%%%%%%%%%%%%%%%%
\begin{table}
    \centering
    \caption{Summary of outcomes for the BNS mergers in the massive DEF theory with APR4 EOS. The first column lists the model name which combines EOS, coupling strength $B$, and baryon mass of each NS in units of $M_\odot$. The second column shows the ADM mass $M_{\rm ADM}$ of each isolated NS. The third column shows the state of pre-merger scalarization with symbols $\times$, $\bigtriangleup$ and $\bigcirc$ corresponding to no scalarization, dynamical scalarization, and spontaneous scalarization in the pre-merger phase, respectively. The fourth column lists the post-merger remnants with LL, SL and PC being a long-lived HMNS, a short-lived HMNS, and prompt collapse. The last two columns summerize the lifetime of the HMNS $\Tns$ and scalar cloud $\Tsc$ for the cases of LL and SL, with '-' representing the absence of descalarization in the post-merger phase.
}
    \begin{tabular}{
        >{\centering}p{0.30\columnwidth} %model name
	>{\centering}p{0.115\columnwidth} %M_ADM
        >{\centering}p{0.150\columnwidth} %pre-merger scalarizaion
	>{\centering}p{0.120\columnwidth} %outcome
        >{\centering}p{0.105\columnwidth} %HMNS life-time
        >{\centering\arraybackslash}p{0.10\columnwidth} %descalarize time
    }
        \hline\hline \\[-.8em]
        Model name
        %& $B$ & $M_{\rm b}$ ($M_\odot$) 
        &  $M_{\rm ADM}$ ($M_\odot$) 
        %& Pre-merger scalarization 
        & Pre-merger $\varphi$
        & Fate 
        & $\Tns$ (ms)
        & $\Tsc$ (ms) \\
        \hline 
        \texttt{APR4\_B13.8\_M1.57} 
            & 1.4040  & $\times$
            & LL & $>10$ & - \\
        \texttt{APR4\_B13.8\_M1.58} 
            & 1.4119  & $\times$
            & PC & 1.13 & - \\
        \hline 
        \texttt{APR4\_B14.3\_M1.57} 
            & 1.4040  & $\times$
            & LL & $>10$ & - \\
        \texttt{APR4\_B14.3\_M1.58} 
            & 1.4119  & $\times$
            & PC & 1.13 & - \\
        \hline 
        \texttt{APR4\_B14.8\_M1.62} 
            & 1.4434  & $\bigtriangleup$
            & SL & 2.13 & 0.52 \\
        \texttt{APR4\_B14.8\_M1.63} 
            & 1.4513  & $\bigtriangleup$
            & PC & 1.17 & 0.50 \\
        \hline 
        \texttt{APR4\_B15.3\_M1.48} 
            & 1.3323  & $\bigtriangleup$
            & LL & $>10$ & - \\
        \texttt{APR4\_B15.3\_M1.50} 
            & 1.3483  & $\bigtriangleup$ 
            & LL & $>10$ & - \\
        \texttt{APR4\_B15.3\_M1.52} 
            & 1.3643  & $\bigcirc$ 
            & LL & $>10$ & 5.28 \\
        \texttt{APR4\_B15.3\_M1.54} 
            %& \multirow{8}{*}{15.3} & 1.54  
            & 1.3802  & $\bigcirc$
            & LL & $>10$ & 4.06 \\
        \texttt{APR4\_B15.3\_M1.56} 
            %& & 1.56  
            & 1.3961  & $\bigcirc$
            & LL & $>10$ & 4.65 \\
        \texttt{APR4\_B15.3\_M1.58} 
            %& & 1.58 
            & 1.4119  & $\bigcirc$
            & LL & $>10$ & 1.66 \\
        \texttt{APR4\_B15.3\_M1.60} 
            %& & 1.60 
            & 1.4276 & $\bigcirc$
            & LL & $>10$ & 1.86 \\
        \texttt{APR4\_B15.3\_M1.62}
            %& & 1.62
            & 1.4433  & $\bigcirc$
            & SL & 2.50 & 0.63 \\
        \texttt{APR4\_B15.3\_M1.64}
            %& & 1.64
            & 1.4590  & $\bigcirc$
            & SL & 2.21 & 0.58 \\
        \texttt{APR4\_B15.3\_M1.65}
            & 1.4668  & $\bigcirc$
            & PC & 1.15 & 0.57 \\
        \texttt{APR4\_B15.3\_M1.66}
            %& & 1.66
            & 1.4746  & $\bigcirc$
            & PC & 1.00 & 0.55 \\
        \texttt{APR4\_B15.3\_M1.68}
            %& & 1.68
            & 1.4901  & $\bigcirc$
            & PC & 0.92 & 0.53 \\
        \hline
        \texttt{APR4\_B15.8\_M1.50}
            & 1.3482  & $\bigcirc$
            & LL & $>10$ & - \\
        \texttt{APR4\_B15.8\_M1.52}
            & 1.3641  & $\bigcirc$
            & LL & $>10$ & - \\
        \texttt{APR4\_B15.8\_M1.54}
            & 1.3800  & $\bigcirc$
            & LL & $>10$ & 6.20 \\
        \texttt{APR4\_B15.8\_M1.56}
            & 1.3959  & $\bigcirc$
            & LL & $>10$ & 5.55 \\
        \texttt{APR4\_B15.8\_M1.58}
            & 1.4116  & $\bigcirc$
            & LL & $>10$ & 4.38 \\
        \texttt{APR4\_B15.8\_M1.60}
            & 1.4274  & $\bigcirc$
            & LL & $>10$ & 1.97 \\
        \texttt{APR4\_B15.8\_M1.62}
            & 1.4430  & $\bigcirc$
            & LL & $>10$ & 1.65 \\
        \texttt{APR4\_B15.8\_M1.64}
            & 1.4586  & $\bigcirc$
            & SL & 2.21 & 1.82 \\
        \texttt{APR4\_B15.8\_M1.65}
            & 1.4664  & $\bigcirc$
            & SL & 3.06 & 0.82 \\
        \texttt{APR4\_B15.8\_M1.66}
            & 1.4742  & $\bigcirc$
            & PC & 1.31 & 0.67 \\
        \texttt{APR4\_B15.8\_M1.67}
            & 1.4820  & $\bigcirc$
            & PC & 1.06 & 0.64 \\
        \texttt{APR4\_B15.8\_M1.68}
            & 1.4897  & $\bigcirc$
            & PC & 0.98 & 0.60 \\
        \texttt{APR4\_B15.8\_M1.70}
            & 1.5052  & $\bigcirc$
            & PC & 0.91 & 0.57 \\
        \hline
        \texttt{APR4\_B16.3\_M1.52}
            & 1.3638  & $\bigcirc$
            & LL & $>10$ & - \\
        \texttt{APR4\_B16.3\_M1.54}
            & 1.3796  & $\bigcirc$
            & LL & $>10$ & - \\
        \texttt{APR4\_B16.3\_M1.56}
            & 1.3954  & $\bigcirc$
            & LL & $>10$ & 6.02 \\
        \texttt{APR4\_B16.3\_M1.58}
            & 1.4112  & $\bigcirc$
            & LL & $>10$ & 5.40 \\
        \texttt{APR4\_B16.3\_M1.60}
            & 1.4269  & $\bigcirc$
            & LL & $>10$ & 4.07 \\
        \texttt{APR4\_B16.3\_M1.62}
            & 1.4425  & $\bigcirc$
            & SL & 5.03 & 2.97 \\
        \texttt{APR4\_B16.3\_M1.64}
            & 1.4581  & $\bigcirc$
            & SL & 3.21 & 1.73 \\
        \texttt{APR4\_B16.3\_M1.66}
            & 1.4736  & $\bigcirc$
            & SL & 2.23 & 1.70 \\
        \texttt{APR4\_B16.3\_M1.68}
            & 1.4891  & $\bigcirc$
            & SL & 1.89 & 0.92 \\
        \texttt{APR4\_B16.3\_M1.69}
            & 1.4968  & $\bigcirc$
            & SL & 1.12 & 0.78 \\
        \texttt{APR4\_B16.3\_M1.70}
            & 1.5045  & $\bigcirc$
            & PC & 1.03 & 0.71 \\
        \hline\hline
    \end{tabular}
    \label{tab:outcomes_APR4}
\end{table}
\begin{table}
    \centering
    \caption{Same as \cref{tab:outcomes_APR4} but for the MPA1 EOS.}
    \begin{tabular}{
        >{\centering}p{0.30\columnwidth} %model name
	>{\centering}p{0.115\columnwidth} %M_ADM
        >{\centering}p{0.150\columnwidth} %pre-merger scalarizaion
	>{\centering}p{0.120\columnwidth} %outcome
        >{\centering}p{0.105\columnwidth} %HMNS life-time
        >{\centering\arraybackslash}p{0.10\columnwidth} %descalarize time
    }
        \hline\hline \\[-.8em]
        Model name
        &  $M_{\rm ADM}$ ($M_\odot$) 
        & Pre-merger $\varphi$
        & Fate 
        & $\Tns$ (ms)
        & $\Tsc$ (ms) \\
        \hline 
        \texttt{MPA1\_B15.0\_M1.78}
            & 1.5831  & $\times$
            & LL & $>10$ & - \\
        \texttt{MPA1\_B15.0\_M1.79}
            & 1.5909  & $\times$
            & PC & 1.18 & - \\
        \hline 
        \texttt{MPA1\_B15.5\_M1.78}
            & 1.5831  & $\times$
            & SL & 3.06 & - \\
        \texttt{MPA1\_B15.5\_M1.79}
            & 1.5909  & $\times$
            & PC & 1.18 & - \\
        \hline 
        \texttt{MPA1\_B16.0\_M1.60}
            & 1.4398  & $\times$
            & LL & $>10$ & - \\
        \texttt{MPA1\_B16.0\_M1.62}
            & 1.4559 & $\times$
            & LL & $>10$ & - \\
        \texttt{MPA1\_B16.0\_M1.64}
            & 1.4719  & $\times$
            & LL & $>10$ & - \\
        \texttt{MPA1\_B16.0\_M1.66}
            & 1.4880  & $\times$
            & LL & $>10$ & - \\
        \texttt{MPA1\_B16.0\_M1.68}
            & 1.5039  & $\times$
            & LL & $>10$ & 9.27 \\
        \texttt{MPA1\_B16.0\_M1.70}
            & 1.5198  & $\times$
            & LL & $>10$ & 2.76 \\
        \texttt{MPA1\_B16.0\_M1.72}
            & 1.5357  & $\bigtriangleup$
            & LL & $>10$ & 3.00 \\
        \texttt{MPA1\_B16.0\_M1.74}
            & 1.5515  & $\bigtriangleup$
            & LL & $>10$ & 2.01 \\
        \texttt{MPA1\_B16.0\_M1.76}
            & 1.5673  & $\bigcirc$
            & LL & $>10$ & 2.64 \\
        \texttt{MPA1\_B16.0\_M1.78}
            & 1.5831  & $\bigcirc$
            & LL & $>10$ & 0.66 \\
        \texttt{MPA1\_B16.0\_M1.80}
            & 1.5987 & $\bigcirc$
            & LL & $>10$ & 0.63 \\
        \texttt{MPA1\_B16.0\_M1.82}
            & 1.6144  & $\bigcirc$
            & LL & $>10$ & 0.61 \\
        \texttt{MPA1\_B16.0\_M1.84}
            & 1.6300  & $\bigcirc$
            & SL & 2.37 & 0.57 \\
        \texttt{MPA1\_B16.0\_M1.85}
            & 1.6377  & $\bigcirc$
            & PC & 1.30 & 0.56 \\
        \texttt{MPA1\_B16.0\_M1.86}
            & 1.6455  & $\bigcirc$
            & PC & 1.15 & 0.55 \\
        \hline 
        \texttt{MPA1\_B16.5\_M1.60}
            & 1.4398  & $\bigtriangleup$
            & LL & $>10$ & - \\
        \texttt{MPA1\_B16.5\_M1.62}
            & 1.4559  & $\bigcirc$
            & LL & $>10$ & - \\
        \texttt{MPA1\_B16.5\_M1.64}
            & 1.4719  & $\bigcirc$
            & LL & $>10$ & - \\
        \texttt{MPA1\_B16.5\_M1.66}
            & 1.4879  & $\bigcirc$
            & LL & $>10$ & - \\
        \texttt{MPA1\_B16.5\_M1.68}
            & 1.5039  & $\bigcirc$
            & LL & $>10$ & - \\
        \texttt{MPA1\_B16.5\_M1.70}
            & 1.5198  & $\bigcirc$
            & LL & $>10$ & 6.38 \\
        \texttt{MPA1\_B16.5\_M1.72}
            & 1.5356  & $\bigcirc$
            & LL & $>10$ & 10.5 \\
        \texttt{MPA1\_B16.5\_M1.74}
            & 1.5514 & $\bigcirc$
            & LL & $>10$ & 5.74 \\
        \texttt{MPA1\_B16.5\_M1.76}
            & 1.5672  & $\bigcirc$
            & LL & $>10$ & 4.41 \\
        \texttt{MPA1\_B16.5\_M1.78}
            & 1.5829  & $\bigcirc$
            & LL & $>10$ & 1.94 \\
        \texttt{MPA1\_B16.5\_M1.80}
            & 1.5986  & $\bigcirc$
            & LL & $>10$ & 1.90 \\
        \texttt{MPA1\_B16.5\_M1.82}
            & 1.6142  & $\bigcirc$
            & SL & 6.57 & 0.76 \\
        \texttt{MPA1\_B16.5\_M1.84}
            & 1.6297  & $\bigcirc$
            & SL & 2.75 & 0.67 \\
        \texttt{MPA1\_B16.5\_M1.86}
            & 1.6452  & $\bigcirc$
            & SL & 2.77 & 0.62 \\
        \texttt{MPA1\_B16.5\_M1.87}
            & 1.6530  & $\bigcirc$
            & PC & 1.27 & 0.61 \\
        \texttt{MPA1\_B16.5\_M1.88}
            & 1.6607  & $\bigcirc$
            & PC & 1.11 & 0.60 \\
        \texttt{MPA1\_B16.5\_M1.90}
            & 1.6761  & $\bigcirc$
            & PC & 0.95 & 0.57 \\
        \hline
        \texttt{MPA1\_B17.0\_M1.70}
            & 1.5195  & $\bigcirc$
            & LL & $>10$ & - \\
        \texttt{MPA1\_B17.0\_M1.72}
            & 1.5353  & $\bigcirc$
            & LL & $>10$ & - \\
        \texttt{MPA1\_B17.0\_M1.74}
            & 1.5511  & $\bigcirc$
            & LL & $>10$ & - \\
        \texttt{MPA1\_B17.0\_M1.76}
            & 1.5668  & $\bigcirc$
            & LL & $>10$ & 5.79 \\
        \texttt{MPA1\_B17.0\_M1.78}
            & 1.5825  & $\bigcirc$
            & LL & $>10$ & 4.47 \\
        \texttt{MPA1\_B17.0\_M1.80}
            & 1.5981  & $\bigcirc$
            & LL & $>10$ & 3.27 \\
        \texttt{MPA1\_B17.0\_M1.82}
            & 1.6137  & $\bigcirc$
            & SL & 5.67 & 1.90 \\
        \texttt{MPA1\_B17.0\_M1.84}
            & 1.6293  & $\bigcirc$
            & SL & 2.50 & 1.90 \\
        \texttt{MPA1\_B17.0\_M1.86}
            & 1.6448  & $\bigcirc$
            & SL & 2.69 & 0.85 \\
        \texttt{MPA1\_B17.0\_M1.87}
            & 1.6525  & $\bigcirc$
            & SL & 2.95 & 0.74 \\
        \texttt{MPA1\_B17.0\_M1.88}
            & 1.6602  & $\bigcirc$
            & PC & 1.36 & 0.69 \\
        \texttt{MPA1\_B17.0\_M1.90}
            & 1.6756  & $\bigcirc$
            & PC & 1.12 & 0.64 \\
        \hline\hline
    \end{tabular}
    \label{tab:outcomes_MPA1}
\end{table}
\begin{table}
    \centering
    \caption{Same as \cref{tab:outcomes_APR4} but for the H4 EOS.}
    \begin{tabular}{
        >{\centering}p{0.30\columnwidth} %model name
	>{\centering}p{0.115\columnwidth} %M_ADM
        >{\centering}p{0.150\columnwidth} %pre-merger scalarizaion
	>{\centering}p{0.120\columnwidth} %outcome
        >{\centering}p{0.105\columnwidth} %HMNS life-time
        >{\centering\arraybackslash}p{0.10\columnwidth} %descalarize time
    }
        \hline\hline \\[-.8em]
        Model name
        &  $M_{\rm ADM}$ ($M_\odot$) 
        & Pre-merger $\varphi$
        & Fate 
        & $\Tns$ (ms)
        & $\Tsc$ (ms) \\
        \hline 
        \texttt{H4\_B15.0\_M1.70}
            & 1.5413  & $\times$
            & SL & 3.69 & - \\
        \texttt{H4\_B15.0\_M1.71}
            & 1.5494  & $\times$
            & PC & 1.37 & - \\
        \hline 
        \texttt{H4\_B15.5\_M1.70}
            & 1.5413  & $\times$
            & SL & 2.59 & - \\
        \texttt{H4\_B15.5\_M1.71}
            & 1.5494  & $\times$
            & PC & 1.37 & - \\
        \hline 
        \texttt{H4\_B16.0\_M1.70}
            & 1.5413  & $\times$
            & SL & 2.68 & - \\
        \texttt{H4\_B16.0\_M1.71}
            & 1.5494  & $\times$
            & PC & 1.37 & - \\
        \hline 
        \texttt{H4\_B16.5\_M1.70}
            & 1.5413  & $\times$
            & SL & 2.84 & - \\
        \texttt{H4\_B16.5\_M1.71}
            & 1.5494  & $\times$
            & PC & 1.36 & - \\
        \hline 
        \texttt{H4\_B17.0\_M1.60}
            & 1.4593  & $\times$
            & LL & $>10$ & - \\
        \texttt{H4\_B17.0\_M1.62}
            & 1.4758  & $\times$
            & SL & 8.54 & 8.40 \\
        \texttt{H4\_B17.0\_M1.64}
            & 1.4923  & $\times$
            & SL & 7.35 & 7.22 \\
        \texttt{H4\_B17.0\_M1.66}
            & 1.5087  & $\times$
            & SL & 8.39 & 8.32 \\
        \texttt{H4\_B17.0\_M1.68}
            & 1.5250  & $\times$
            & SL & 2.43 & 2.30 \\
        \texttt{H4\_B17.0\_M1.70}
            & 1.5413  & $\times$
            & SL & 3.94 & 3.84 \\
        \texttt{H4\_B17.0\_M1.72}
            & 1.5576  & $\times$
            & SL & 2.24 & 2.13 \\
        \texttt{H4\_B17.0\_M1.74}
            & 1.5738  & $\bigtriangleup$
            & SL & 1.52 & 1.56 \\
        \texttt{H4\_B17.0\_M1.75}
            & 1.5819 & $\bigtriangleup$
            & PC & 1.18 & 1.12 \\
        \texttt{H4\_B17.0\_M1.76}
            & 1.5899  & $\bigcirc$
            & PC & 1.00 & 1.06 \\
        \texttt{H4\_B17.0\_M1.78}
            & 1.6060  & $\bigcirc$
            & PC & 0.92 & 0.95 \\
        \texttt{H4\_B17.0\_M1.80}
            & 1.6221  & $\bigcirc$
            & PC & 0.83 & 0.87 \\
        \hline 
        \texttt{H4\_B17.5\_M1.60}
            & 1.4593  & $\times$
            & LL & $>10$ & - \\
        \texttt{H4\_B17.5\_M1.62}
            & 1.4758  & $\times$
            & LL & $>10$ & - \\
        \texttt{H4\_B17.5\_M1.64}
            & 1.4923  & $\times$
            & LL & $>10$ & - \\
        \texttt{H4\_B17.5\_M1.66}
            & 1.5087  & $\bigtriangleup$
            & SL & 6.63 & 6.53 \\
        \texttt{H4\_B17.5\_M1.68}
            & 1.5250  & $\bigcirc$
            & SL & 7.00 & 7.95 \\
        \texttt{H4\_B17.5\_M1.70}
            & 1.5413  & $\bigcirc$
            & SL & 4.66 & 4.62 \\
        \texttt{H4\_B17.5\_M1.72}
            & 1.5575  & $\bigcirc$
            & SL & 2.27 & 2.24 \\
        \texttt{H4\_B17.5\_M1.74}
            & 1.5737  & $\bigcirc$
            & SL & 2.11 & 1.95 \\
        \texttt{H4\_B17.5\_M1.75}
            & 1.5817 & $\bigcirc$
            & SL & 2.16 & 2.01 \\
        \texttt{H4\_B17.5\_M1.76}
            & 1.5898  & $\bigcirc$
            & PC & 1.27 & 1.16 \\
        \texttt{H4\_B17.5\_M1.78}
            & 1.6058  & $\bigcirc$
            & PC & 1.00 & 1.03 \\
        \texttt{H4\_B17.5\_M1.80}
            & 1.6218 & $\bigcirc$
            & PC & 0.90 & 0.96 \\
        \texttt{H4\_B17.5\_M1.82}
            & 1.6377  & $\bigcirc$
            & PC & 0.87 & 0.90 \\
        \texttt{H4\_B17.5\_M1.84}
            & 1.6535 & $\bigcirc$
            & PC & 0.78 & 0.83 \\
        \texttt{H4\_B17.5\_M1.86}
            & 1.6693  & $\bigcirc$
            & PC & 0.73 & 0.81 \\
        \hline 
        \texttt{H4\_B18.0\_M1.60}
            & 1.4593  & $\bigcirc$
            & LL & $>10$ & - \\
        \texttt{H4\_B18.0\_M1.62}
            & 1.4758  & $\bigcirc$
            & LL & $>10$ & - \\
        \texttt{H4\_B18.0\_M1.64}
            & 1.4922  & $\bigcirc$
            & LL & $>10$ & - \\
        \texttt{H4\_B18.0\_M1.66}
            & 1.5086  & $\bigcirc$
            & LL & $>10$ & 9.96 \\
        \texttt{H4\_B18.0\_M1.68}
            & 1.5248  & $\bigcirc$
            & SL & 7.63 & 7.59 \\
        \texttt{H4\_B18.0\_M1.70}
            & 1.5410  & $\bigcirc$
            & SL & 5.82 & 5.79 \\
        \texttt{H4\_B18.0\_M1.72}
            & 1.5572  & $\bigcirc$
            & SL & 2.29 & 2.22 \\
        \texttt{H4\_B18.0\_M1.74}
            & 1.5733  & $\bigcirc$
            & SL & 2.11 & 2.14 \\
        \texttt{H4\_B18.0\_M1.76}
            & 1.5893  & $\bigcirc$
            & SL & 2.02 & 1.95 \\
        \texttt{H4\_B18.0\_M1.77}
            & 1.5973  & $\bigcirc$
            & SL & 2.55 & 2.49 \\
        \texttt{H4\_B18.0\_M1.78}
            & 1.6053  & $\bigcirc$
            & PC & 1.16 & 1.10 \\
        \texttt{H4\_B18.0\_M1.80}
            & 1.6212 & $\bigcirc$
            & PC & 1.01 & 1.02 \\
        \texttt{H4\_B18.0\_M1.82}
            & 1.6371  & $\bigcirc$
            & PC & 0.87 & 0.95 \\
        \texttt{H4\_B18.0\_M1.84}
            & 1.6528 & $\bigcirc$
            & PC & 0.81 & 0.91 \\
        \texttt{H4\_B18.0\_M1.86}
            & 1.6686  & $\bigcirc$
            & PC & 0.82 & 0.86 \\
        \hline\hline
    \end{tabular}
    \label{tab:outcomes_H4}
\end{table}
%%%%%%%%%%%%%%%%%%%%%%%%%%%%%%%%%%%%%%%%%%%%%%%%%%%%%%%

%%%%%%%%%%%%%%%%%%%%%%%%%%%%%%%%%%%%%%%%
\bibliographystyle{apsrev4-2}
\bibliography{references}

%%%%%%%%%%%%%%%%%%%%%%%%%%%%%%%%%%%%%%%%

\end{document}